\documentclass[pdflatex,sn-mathphys-num, iicol]{sn-jnl}% Math and Physical Sciences Numbered Reference Style
\usepackage{braket,ulem}
\usepackage[utf8]{inputenc}
\usepackage{graphicx}%
\usepackage{multirow}%
\usepackage{amsmath,amssymb,amsfonts}%
\usepackage{amsthm}%
\usepackage{mathrsfs}%
\usepackage[title]{appendix}%
\usepackage{xcolor}%
\usepackage{textcomp}%
\usepackage{manyfoot}%
\usepackage{booktabs}%
\usepackage{algorithm}%
\usepackage{algorithmicx}%
\usepackage{algpseudocode}%
\usepackage{listings}%
\usepackage{comment}
\usepackage{hyperref}
\usepackage{url}
\usepackage{xurl} 
\usepackage{physics}
\newcommand\MAD[1]{{\color{red} [MAD] }}

\newcommand{\aspas}[1]{``#1''}
\theoremstyle{thmstyleone}%
\theoremstyle{thmstyletwo}%

\theoremstyle{thmstylethree}%

\begin{document}

\title[Article Title]{Towards a point-to-point CV-QKD system: Implementation challenges and perspectives}

%\title[Article Title]{Towards an End-to-End CV-QKD System: Physical Challenges and Implementation Perspectives for Brazil's Quantum Communication Infrastructure}


\author*[1]{\fnm{Davi} \sur{Juvêncio Gomes de Sousa}}
\email{davi.juvencio@fieb.org.br}

\author[1]{\fnm{Nelson} \sur{Alves Ferreira Neto}}
\email{nelson.neto@fieb.org.br}

\author[1]{\fnm{Christiano} \sur{M. S. Nascimento}}
\email{christiano.moreira@fieb.org.br}

\author[1]{\fnm{Lucas} \sur{Q. Galvão}}
\email{lqgalvao3@gmail.com}

\author[1]{\fnm{Mauro} \sur{Queiroz Nooblath Neto}}
\email{mauro.neto@fieb.org.br}

\author[2,1]{\fnm{Micael} \sur{Andrade Dias}}\email{mandi@dtu.dk}

\author[1]{\fnm{Cássio} \sur{de Castro Silva}}\email{cassio.castro@fbter.org.br}

\author[1]{\fnm{Braian} \sur{Pinheiro da Silva}}\email{braian.silva@fieb.org.br}

\author[1]{\fnm{Alexandre} \sur{B. Tacla}}
\email{alexandre.tacla@fieb.org.br}

\author[1]{\fnm{Valéria} \sur{Loureiro da Silva}}
\email{valeria.dasilva@fieb.org.br}

\affil*[1]{\orgdiv{QuIIN – Quantum Industrial Innovation, EMBRAPII CIMATEC Competence Center in Quantum Technologies}, \orgname{SENAI CIMATEC}, \orgaddress{\street{Av. Orlando Gomes 1845}, \city{Salvador}, \postcode{41650-010}, \state{BA}, \country{Brasil}}}

\affil[2]{\orgdiv{Department of Electrical and Photonics Engineering}, \orgname{Technical University of Denmark}, \orgaddress{\street{Ørsteds Plads, Building 340}, \city{Lyngby}, \postcode{2800}, \country{Denmark}}}


\abstract{

This article presents an analysis of the practical challenges and implementation perspectives of point-to-point continuous-variable quantum key distribution (CV-QKD) systems over optical fiber. The study addresses the physical layer, including the design of transmitters, quantum channels, and receivers, with emphasis on impairments such as attenuation, chromatic dispersion, polarization fluctuations, and coexistence with classical channels. We further examine the role of digital signal processing (DSP) as the bridge between quantum state transmission and classical post-processing, highlighting its impact on excess noise mitigation, covariance matrix estimation, and reconciliation efficiency. The post-processing pipeline is detailed with a focus on parameter estimation in the finite-size regime, information reconciliation using LDPC-based codes optimized for low-SNR conditions, and privacy amplification employing large-block universal hashing. From a hardware perspective, we discuss modular digital architectures that integrate dedicated accelerators with programmable processors, supported by a reference software framework (\textit{CV-QKD-ModSim}) for algorithm validation and hardware co-design. Finally, we outline perspectives for the deployment of CV-QKD in Brazil, starting from metropolitan testbeds and extending toward hybrid fiber/FSO and space-based infrastructures. The work establishes the foundations for the first point-to-point CV-QKD system in Brazil, while providing a roadmap for scalable and interoperable quantum communication networks.
}




\maketitle

\section{Introduction}\label{sec1}

A century has passed since the foundational works of quantum mechanics revolutionized our understanding of nature at its most fundamental level. Since then, our understanding of the quantum world has evolved from a purely physical perspective to an informational one. The emergence of quantum information science has led to the second quantum revolution, transforming our view of quantum mechanics from a purely physical theory to a powerful informational resource~\cite{deutsch20}. This paradigm shift has laid the groundwork for quantum technologies that are reshaping our modern world. 

Today, as we commemorate these 100 years of quantum physics, Brazil finds itself at a pivotal moment in the quantum technology landscape, with quantum communication emerging as one of the most promising and strategically important applications of quantum mechanics. Recent large-scale initiatives in quantum communication and cryptography have been actively promoted by Brazil's Ministry of Science, Technology and Innovation (MCTI), resulting in the current development of quantum networks in Recife (PE), Rio de Janeiro (RJ), and São Carlos (SP)~\cite{noticia_fapesp_RRQ_etc, rederio}. Furthermore, the EMBRAPII CIMATEC Competence Center in Quantum Technologies, called Quantum Industrial Innovation (QuIIN), seeks to position Brazil at the forefront of scientific and technological development of quantum technologies. 

Established in December 2023 through a partnership between SENAI CIMATEC, MCTI, and EMBRAPII (the Brazilian Company for Industrial Research and Innovation), QuIIN's mission is to advance quantum technologies in Brazil by fostering collaborative research and development partnerships with academia and industry, providing training and capacity building in quantum technologies (mainly in quantum communication and quantum computing), and supporting the creation and development of startups in the quantum sector. Despite being a young initiative, QuIIN already has a multidisciplinary team of over 80 researchers (staff and fellows), including physicists, mathematicians, computer scientists, and engineers. With a strong focus on continuous-variable quantum key distribution (CV-QKD), the research team is tackling a range of theoretical and experimental challenges, including hardware development, with the ambitious goal of implementing Brazil's first point-to-point CV-QKD system. Being the first initiative of this nature in the country adds a layer of complexity, as it requires not only overcoming the inherent technical difficulties of CV-QKD, but also building local expertise, establishing experimental infrastructure from the ground up, and navigating uncharted regulatory and integration pathways. In this pioneering effort, the team must simultaneously act as developers, educators, and ecosystem builders.

Quantum Key Distribution (QKD) is one of the first quantum technologies to achieve practical implementation at scale. Over the past two decades, the field has evolved from theoretical proposals to commercial solutions and advanced experimental demonstrations~\cite{Pirandola_2020, zhang2024continuous}. Thanks to continuous progress in protocol design, rigorous security analyses, and technological development, large-scale testbed deployments are now operating worldwide, ranging from metropolitan networks~\cite{chen2025implementation,Cao2022,wei2022towards,liu2025road} to intercontinental satellite-based quantum communication links~\cite{Liao2017, Yin2017, Ren2017, li2025microsatellite,Bedington2017, chen2025implementation}.

Although discrete-variable QKD (DV-QKD) protocols were the first to be developed~\cite{Bennett1984} and achieve widespread adoption~\cite{Pirandola_2020}, in recent years CV-QKD systems have emerged as a competitive alternative, advancing rapidly in both theoretical understanding and experimental implementations~\cite{zhang2024continuous}. The fundamental distinction between these approaches lies in the quantum states employed as information carriers and their corresponding measurement strategies. In a typical DV-QKD protocol, the transmitter (usually called Alice) encodes classical bits of information in discrete degrees of freedom of single photons (e.g., horizontal/vertical polarization) and transmits them through a quantum channel to the receiver (called Bob), who measures each photon to obtain binary outcomes. In most CV-QKD protocols, however, information is encoded in the quadratures of coherent states generated by laser sources; Alice transmits states modulated according to a predetermined encoding scheme, where different values of quadratures correspond to different information symbols. After receiving the CV state, Bob measures the quadratures by performing homodyne (or heterodyne) detection, thus obtaining continuous-valued measurement results. 

 Because classical optical communication systems also encode information in the quadrature amplitudes of laser light and recover it through coherent detection, CV-QKD protocols are fully compatible with the same fundamental components that form the backbone of modern telecommunication networks. This compatibility enables higher key generation rates over short and medium distances (of the order of Gbps for $\sim$10~km and Mbps for $\sim$100~km~\cite{wang2025, hajomer2024}), facilitates integration into photonic platforms~\cite{hajomer2024b,hajomer2025chipbased} and existing fiber networks~\cite{hajomer2025coexistence}. This approach thus provides a cost-effective alternative that eliminates the need for new infrastructure with specialized single-photon detection equipment, which is required by DV systems~\cite{zhang2024continuous}. 

However, these advantages come at the cost of significantly increased complexity in post-processing and digital signal processing (DSP), which translates into increased hardware complexity compared to DV-QKD systems. Unlike discrete-variable systems, where single photon detection events provide relatively clean binary outcomes, CV-QKD must extract cryptographic keys from continuous measurement data that are inherently dominated by quantum and classical noise. This fundamental challenge requires sophisticated DSP algorithms to distinguish and recover the highly attenuated quantum signal from background noise, demanding real-time computational capabilities with high precision analog-to-digital conversion and extensive digital filtering. Furthermore, the continuous nature of the measurement data necessitates parameter estimation procedures that can accurately characterize the quantum channel and assess security parameters. The post-processing pipeline must implement advanced error-correction codes specifically designed to operate at very low signal-to-noise ratios (SNRs) in continuous-variable systems. In practice, information reconciliation is typically performed using multi-edge low-density parity-check (LDPC) codes acting on blocks of at least $10^{6}$ symbols to enable reliable decoding at low code rates. In contrast, parameter estimation and privacy amplification require significantly larger block sizes, ideally $10^{8}$ symbols or more, in order to mitigate finite-size effects and approach the asymptotic secret key rate. In a point-to-point CV-QKD system, all these operations must be executed in real time, transforming what might initially seem like a simpler optical setup into a complex signal processing and systems engineering challenge. This high level of computational and algorithmic sophistication, combined with the need for robust and scalable hardware, makes the CV-QKD implementation a non-trivial task. 

In the Brazilian context, these challenges are magnified by the absence of existing quantum communication infrastructure, which demands not only technological innovation but also the creation of a national knowledge base and ecosystem around quantum communication. Against this backdrop, this article explores the technical, infrastructural, and strategic challenges of developing Brazil's first point-to-point CV-QKD system, and outlines the perspectives that such an effort opens for the country’s role in the global quantum technology landscape. To complement this practical perspective, another article in this special issue presents an introductory review of CV-QKD theory~\cite{anka2025}, covering fundamental concepts and key protocols in the field. For more comprehensive treatments of the subject, readers are referred to~\cite{laudenbach2018continuous, djordjevic2019physical, weedbrook2012gaussian, Pirandola_2020, usenko2025continuousvariablequantumcommunication,Diamanti2015}. Additionally, to support the training of a new generation of Brazilian researchers specializing in QKD, an accessible tutorial on quantum cryptography in Portuguese is presented in~\cite{sena2025}.

This article is organized as follows. In Sec.~\ref{sec:2_theory}, we provide an overview of Quantum Key Distribution (QKD) protocols, with a focus on Continuous Variable QKD (CV-QKD) systems. In Sec.~\ref{sec: System}, we describe the physical layer of CV-QKD implementations, detailing the transmitter, channel, and receiver subsystems. In Sec.~\ref{sec:4_post-proc}, we present the post-processing pipeline, including digital signal processing, parameter estimation, information reconciliation, and privacy amplification. Sec.~\ref{sec:hardware_development} discusses the hardware development efforts required to enable real-time CV-QKD, including digital architectures and reference simulation frameworks. Finally, Sec.~\ref{sec:perspectives} presents perspectives for the deployment of quantum communication infrastructure in Brazil, with emphasis on ongoing initiatives and future scalability.

\section{CV-QKD Systems Overview}
\label{sec:2_theory}

A quantum key distribution (QKD) protocol is a quantum communication system designed to leverage unique properties of quantum systems to share classical random sequences secretly. The objective of a QKD protocol is to allow the two legitimate parties, Alice and Bob (the transmitter and receiver, respectively), to distribute (or generate) a classical random sequence that is kept secret from a potential eavesdropper, named Eve. This classical random sequence, the key, can then be used in further cryptographic protocols, such as one-time-pad (OTP) \cite{shannon1949}, the Advanced Encryption Standard (AES) \cite{manimozhi2025}, or key expansion protocols \cite{zhang2022}.

\begin{figure}[!tb]
    \centering
    \includegraphics[width=\linewidth, page=2]{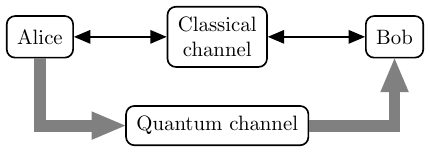}
    \caption{Conceptual model of a QKD protocol. }
    \label{fig:QKD_model}
\end{figure}

In a conceptual picture, a QKD protocol makes use of a noisy insecure quantum channel and a public authenticated classical channel, as depicted in Fig. \ref{fig:QKD_model}. The quantum channel is used for the quantum communication stage and it is considered to be under Eve's control, meaning that she can interact with the quantum states sent by Alice to gain information on the shared key. As a security premise, the noise coming from the quantum channel is assumed to be the result of an eavesdropping attempt \cite{usenko2016}. The classical channel, on the other hand, is used by Alice and Bob to exchange messages during the classical postprocessing of the raw key. By hypothesis, the classical channel is public, meaning that the classical messages are publicly available, but it is authenticated, so that Alice and Bob are certain that they are receiving messages from each other and not by the eavesdropper in disguise.

In a CV-QKD protocol, coherent states of light are typically used since they are easily generated in the laboratory by telecom-grade lasers, and information is encoded on conjugated quadratures of the electromagnetic field \cite{Diamanti2015}. The quadratures may be modulated according to a pair of independent Gaussian random variables, or a discretized version with an appropriate choice of a discrete probability distribution on a set of points on the phase space (the complex plane) \cite{grosshans2002,Weedbrook2004WithoutSwitching,ghorai2019,djordjevic2019,denys2021}. Protocols following the first strategy compose the class of Gaussian modulated coherent states (GMCS) protocols, whilst the second modulation strategy is known as discrete modulation of CV-QKD protocols. At the receiver, coherent measurement is employed by Bob, which can be either a homodyne detector, with a random choice of which quadrature is to be measured, or a heterodyne detector that measures both quadratures at once.

The operation of a typical CV-QKD protocol comes in general four major stages, (i) quantum communication, (ii) parameter estimation, (iii) information reconciliation, and (iv) privacy amplification. Below, we outline the important case of the GG02 protocol~\cite{grosshans2002}, which employs Gaussian-modulated coherent states with homodyne detection. A variant of this protocol with heterodyne (double homodyne) detection was later introduced by Weedbrook \textit{et al.} in 2004~\cite{Weedbrook2004WithoutSwitching}. Both versions produce Gaussian random variables and can be analyzed using similar security frameworks~\cite{Pirandola_2020}, though they differ in their implementations and measurement noise characteristics. Heterodyne detection, which measures both quadratures simultaneously, has become more common in practice as it eliminates the need for a quantum random number generator (QRNG) at the receiver and the sifting stage. However, homodyne detection introduces less noise by measuring only one quadrature at a time, potentially enabling higher secret key rates.

\begin{enumerate}
    \item \textbf{Quantum communication} - Comprises the \aspas{quantum part} of a CV-QKD protocol where quantum states are prepared, transmitted and measured. It consists of the following steps that are repeated $L$ times:
    \begin{enumerate}
        \item \textit{State preparation} - Alice has access to a pair of independent and identically Gaussian random variables $Q,P\sim\mathcal{N}(0,\Tilde{V}_m)$. At each round, Alice draws independent samples from $P$ and $Q$, which she uses to prepare the coherent state $\ket{\alpha} = \ket{q + ip}$ by modulating the quadratures of a coherent pulse. The random samples of both $Q$ and $P$ are stored on the registers $Q_L$ and $P_L$.
        
        \item \textit{Quantum state transmission} - Alice transmits the modulated quantum signal through the insecure quantum channel, which is typically modeled by the transmittance parameter $\tau$, accounting for the channel attenuation, and the excess noise $\xi$. Usually, implementations of CV-QKD protocols transmit the coherent states through optical fibers or free-space. During propagation, the quantum signals experience multiple distortion processes, which are discussed in detail in Section~\ref{sec: System}.
        \item \textit{Measurement} - At reception, Bob must randomly chose to measure either the $q$ or $p$ quadrature at each round, and applies the homodyne measurement. The measurement results are stored in the $L$-sized register $Y_L$.
    \end{enumerate}
    
    \item \textbf{Parameter estimation} - Alice and Bob have to estimate the channel parameters allowing them to compute the secret key rate (SKR). Before it takes place, Bob must inform Alice the random quadrature choices of the homodyne detection. Alice then discards the values corresponding to non-measured quadratures and form the random vector $X_L$. This procedure is called key sifting\footnote{It is the CV analogous of the sifting procedure used in the BB84 DV-QKD protocol~\cite{Bennett1984}.} and is only required in protocols with homodyne detection, where only one quadrature is measured. The pair $X_L$ and $Y_L$ is called the shared raw key. Now, Alice and Bob perform an estimation of the security parameters of the system by choosing a random subset of size $L'\lll L$ to be announced publicly and used by the estimators. The samples used for parameter estimation are discarded so that Alice and Bob keep the remaining $l = L - L'$ raw key elements. The estimated parameters are used to compute the SKR for the exchanged states. The protocol aborts if the SKR is negative, indicating that the channel conditions do not allow for secure key extraction. Otherwise, the protocol proceeds to the information reconciliation stage.
    
    \item \textbf{Information reconciliation} - After parameter estimation, the sequences $X_l$ and $Y_l$ are correlated random variables that must be used to distill a pair of identical binary sequences. This process is accomplished with the use of an Information Reconciliation protocol, which will be discussed in detail in Section \ref{sec:information-reconciliation}. The general idea is that, as $X_l$ and $Y_l$ are Gaussian vectors, a quantization operation\footnote{Here, quantization refers to the representation of a continuous-valued variable with finite precision. Not to be confused with the quantization of the electromagnetic field.} must take place so that the generated key takes binary values, followed by an error correction protocol to ensure that there are no differences between Alice and Bob's sequences. We denote by $Q(\cdot)$ the quantization operation. Alice and Bob may chose between performing direct reconciliation (DR) or reverse reconciliation (RR). In the DR scenario, the reference frame is $U_A = Q(X_l)$ and Bob applies error correction to $Y_l$, obtaining a binary sequence $U_B$ such that $U_A = U_B$ with high probability. In the RR, the roles are reversed and the reference frame is $U_l = Q(Y_l)$. Counterintuitively, RR is preferred as it allows key distribution beyond the 3 dB loss limit \cite{assche2004}.
    
    \item \textbf{Privacy amplification} - After information reconciliation, Alice and Bob share a pair of binary sequences $(U_A,U_B)$ that are equal with high probability, but insecure in the sense that Eve acquired information during quantum state transmission and also from the reconciliation messages. To remove Eve's knowledge, Alice and Bob must reduce the size of the shared key by a fraction that is determined by the SKR calculation in the parameter estimation step. This reduction should enforce both randomness and secrecy, and is accomplished by using a suitable random mapping $f: \qty{0,1}^l\mapsto\qty{0,1}^m$, where $m$ is the final key length given by security analysis. The mapping is randomly chosen from a family of 2-universal hash functions ensuring that $S=f(U_A) = f(U_B)$ with high probability, which is the final key, and that Eve has no information on $S$.
\end{enumerate}

The above description provides a general outline of CV-QKD protocols. A few details deserve particular attention in practical implementations. The quantum channel parameters are unknown and must be estimated under the conservative assumption that the channel is controlled by an eavesdropper. Since information is encoded in coherent state quadratures, the channel naturally introduces attenuation and excess noise beyond the shot noise limit, making parameter estimation crucial for security analysis and SKR computation, as will be discussed in Section \ref{sec:parameter_estimation}. Historically, parameter estimation was performed before error correction, but reversing this order has proven more efficient. This enables the use of nearly all exchanged data for both tasks, leading to more accurate channel estimation and effectively enlarging the dataset available for secret key distillation~\cite{hajomer2024, jain2022, leverrier2015}. This is possible because information reconciliation depends on absolute SNR, which can be estimated without disclosing raw data, while specific security related channel parameters are only available though by data disclosure. A generalization to a protocol using heterodyne detection can be done with adjustments on the raw key length to be $2L$ and by dropping the sifting procedure \cite{Weedbrook2004WithoutSwitching}. Information reconciliation and privacy amplification should proceed equally in this case.

Another important aspect of CV-QKD protocols protocols is the reconciliation direction. In DR there is a fundamental 3 dB loss limit. Beyond this threshold, the channel attenuation is so severe that an eavesdropper on the channel would receive more information than Bob, making secure key distillation impossible. This occurs because, in a lossy channel, the eavesdropper can in principle capture the lost light, and beyond 3 dB loss (50\% transmittance), more information is lost to the environment than reaches Bob. While techniques such as advantage distillation were proposed to overcome this limit~\cite{silberhorn2002}, they lack compatibility with broader security proof frameworks. In contrast, RR does not face this limitation. Since Bob holds the reference frame, secure keys can theoretically be generated at any distance, with excess channel noise--rather than loss--becoming the primary limiting factor.

The above description of a CV-QKD protocol covers the major practical tasks that are are required by a theoretical security proof, meaning that it is, in some sense, a conceptual description. A practical implementation can result in several other aspects not foreseen in the theoretical model, which can impact the protocol performance by a significant increase on the overall noise levels, or even by introducing security loopholes due to specific system architectures or devices vulnerabilities \cite{yang2023information,chen2023continuous}.

Due to the variety of practical implementation possibilities, several CV-QKD architectures have been proposed, with certain key innovations serving as major milestones in the development of the technique. The characteristics of transmitter and receiver setups will be detailed in the next sections. Notable advances include the transition from transmitted local oscillator (TLO) to local local oscillator (LLO) schemes.
In TLO, Alice transmits her local oscillator alongside the quantum signal to Bob, while in LLO, Bob generates it locally and corrects the phase using a reference signal from Alice \cite{soh2015}. Additional key developments are the increasing use of digital signal processing \cite{roumestan2021high,pereira2022,hajomer2024} and the refinement of error correction codes (ECC), especially decoding implementations, as part of the information reconciliation protocol \cite{milicevic2018quasi,bai2017}.

The transition from TLO to LLO is due to security reasons. While using TLO is more practical in the sense that the transmitted local oscillator serves directly as the phase reference for coherent detection, it results in a security loophole as the eavesdropper can manipulate the reference signal and change Bob’s measurements. The LLO implementation solves this security problem by avoiding the transmission of a strong reference field, while adding complexity to the receiver as it must synchronize both lasers (Alice and Bob’s sides), typically through digital signal processing techniques. Digital signal processing has become imperative as it allows Bob to compensate for the optical channel effects dispersions and unavoidable devices imperfections. The efficient implementations of long- and low-rate decoders for ECC's, especially for LDPC codes, allows the system to operate at very low signal to noise region, meaning longer distances.

\section{Physical Layer}
\label{sec: System}

QKD systems are primarily based on optical and photonic components. While information can be encoded in various optical degrees of freedom--such as polarization, time-bin, or spatial modes--CV-QKD encodes information in the field quadratures of light. This approach offers natural compatibility with conventional optical telecommunication technologies, making CV-QKD particularly suitable for integration into existing fiber-optic networks. In this section, we briefly describe the optical implementation of a typical CV-QKD systems, including the key components for state preparation, transmission, and detection.

\subsection{Transmitter}
% descrever o bloco do transmissor

\begin{figure*}[!h]
    \centering
    \includegraphics[width=0.7\textwidth]{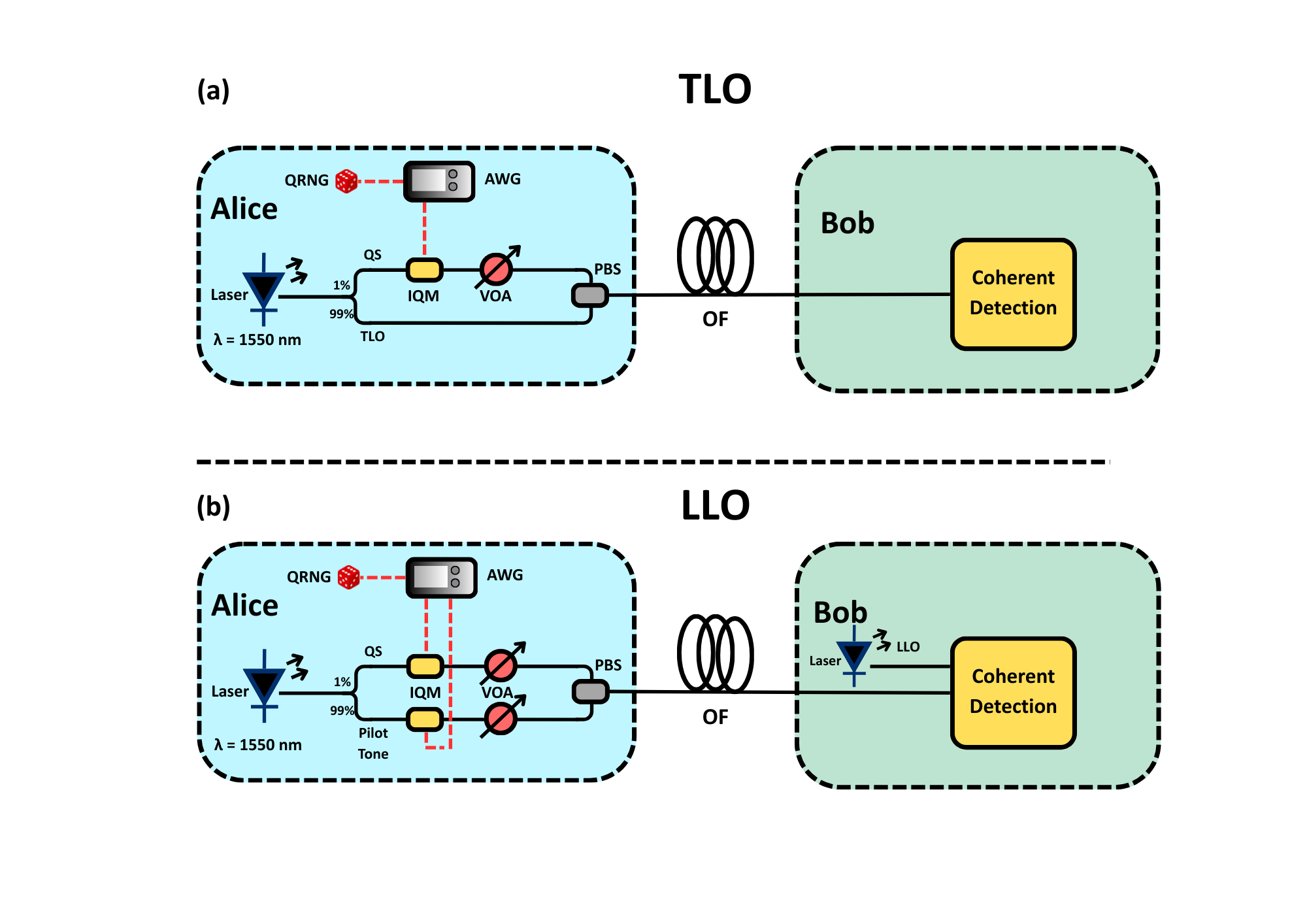}
    \caption{Different transmission systems. (a) Transmitting Local Oscillator (TLO), in which the local oscillator is generated at Alice and co-propagates with the quantum signal through the optical fiber. A Telecom laser ($\lambda$ = 1550nm) is split such that a fraction is called quantum signal (QS), which is modulated by an I/Q modulator (IQM) driven by an arbitrary waveform generator (AWG) seeded by a quantum random number generator (QRNG). The remaining optical power is transmitted to function as a local oscillator (TLO). A variable optical attenuator (VOA) sets the appropriate signal power before transmission, and a polarization beam splitter (PBS) is employed to multiplex the optical fields. At the receiver, Bob performs coherent detection, which will be explained in Sec. (b) Local-local Oscillator (LLO) configuration, where the quantum signal and a pilot tone are generated at Alice, while the local oscillator is independently generated at Bob. The pilot tone is transmitted along with the quantum signal to enable phase and frequency recovery at the receiver. At Bob, a local laser acts as the LLO and is combined with the received signal for coherent detection.}
    \label{fig:3_LLO_TLO}
\end{figure*}

Typical configurations of a CV-QKD optical system are depicted in Fig. \ref{fig:3_LLO_TLO}: (a) with a TLO and (b) with a LLO. The system consists of three main parts: the transmitter (Alice), the quantum channel, and the receiver (Bob). As discussed in Sec. \ref{sec:2_theory}, Alice prepares and transmits a random coherent state to Bob in a typical prepare-and-measure CV-QKD protocol. Alice's transmitter is based on a standard commercial laser, which generates strong coherent states. These states are modulated using an in-phase and quadrature (IQ) electro-optic modulator to encode Gaussian-distributed random variables onto the field quadratures. A variable optical attenuator (VOA) then reduces the signal to the quantum regime, with a mean photon number on the order of one photon per pulse~\cite{zhang2024continuous}. While coherent states remain the primary choice for practical CV-QKD systems, squeezed states of light have been theoretically shown to offer advantages such as enhanced secret key rates, improved resilience to excess noise, and better performance under imperfect information reconciliation~\cite{usenko2018unidimensional,zhang2024continuous}. Despite these benefits, experimental implementations face significant technical challenges. Recent work has demonstrated the practical viability of squeezed-state CV-QKD over optical fiber channels, moving beyond earlier free-space demonstrations with emulated loss \cite{nguyen2025, nguyen2025practical,nag2025continuous}.

% \begin{figure*}[!h]
%     \centering
%     \includegraphics[width=0.7\textwidth]{images/Optical_system.png}
%     \caption{Optical System for a CV-QKD implementation. (a) - The three sub-parts of a system, namely, Alice, Bob and the channel. IQM - IQ modulator, OF - Optical fiber. (b) - Implementation of an attenuated coherent state (VOA - Variable optical attenuator). (c) - Implementation of a Squeezed state.}
%     \label{fig:3_Optical_system}
% \end{figure*}

The choice of laser source is a critical design decision in CV-QKD systems, as the system's security and performance are directly affected by excess noise, with phase noise being a dominant contributor. Phase noise is determined by the laser's linewidth: a broader linewidth introduces larger phase fluctuations, degrading the SNR and reducing the achievable secret key rate. Although many demonstrations employ narrow linewidths (around 100 Hz~\cite{hajomer2024,hajomer2025coexistence}) to mitigate these issues, such sources are costly. For scalable CV-QKD network deployment, this cost factor becomes a significant practical constraint. Consequently, the choice of laser is not merely a hardware consideration but a strategic design decision: selecting a laser with sufficiently low phase noise and with viable cost is paramount to ensuring both the system's performance and practical feasibility.

As mentioned, information is encoded in the quadratures of the laser field. This is commonly achieved using optical modulators, which exploit the Kerr effect to convert electrical signals into optical modulation. Although alternative encoding strategies exist, such as polarization encoding implemented with Pockels cells, these approaches are generally more complex to realize experimentally. Consequently, optical modulators remain the predominant solution. The first experimental implementation employed Gaussian modulation \cite{gg02}. More recently, discrete modulation formats, such as QAM, have attracted increasing interest \cite{zhang2024continuous}.

%Regardless of which state is being used, the information is on the quadratures of light. One way is by implementing optical modulators, which, through the Kerr effect, convert electrical signals into optical signals. Other ways of coding information are more complex to use, such as the use of a Pockels' cell for polarization; therefore, the use of optical modulators is mainly the primary option. The first implementation of a CV-QKD protocol utilized Gaussian modulation; however, over the years, the use of discrete modulation, such as QAMs, has gained more traction \cite{zhang2024continuous}.

%Furthermore, CV-QKD systems are similar to coherent communication systems. The quantum state should maintain amplitude and phase information. So, the necessity of a phase reference is paramount.

Furthermore, CV-QKD systems share similarities with coherent classical communication, as the quantum states must preserve both amplitude and phase information. As a result, the establishment of a reliable phase reference is essential. In early implementations, most protocols relied on TLO to the receiver. However, several attacks targeting this approach have been proposed \cite{qi2015generating,jouguet2013preventing}. The two implementation strategies, TLO and LLO, are illustrated in Fig. \ref{fig:3_LLO_TLO}.

\subsection{Channels}

Over the years, CV-QKD has been predominantly implemented using optical fiber (OFs) channels \cite{ghalaii2023continuous}. Optical fibers offer advantages compared to other transmission media, including high bandwidth, compact deployment, and immunity to electromagnetic interference. In addition, they can be modeled as a Gaussian channel, as discussed in Sec.~\ref{sec:4_post-proc}. Nevertheless, optical fibers are subject to several impairments, and a proper understanding of these limitations is essential for the design of practical systems.

One of the primary impairments is fiber attenuation, which causes the signal transmittance to decrease exponentially with distance. While this loss is acceptable for metropolitan-scale links, it significantly constrains transmission over longer distances. At 1550~nm, which corresponds to a low-loss window, the attenuation of standard fibers is approximately 0.2~dB/km, thereby limiting the achievable secret key rate.

Another relevant limitation is chromatic dispersion, which leads to pulse broadening as a function of wavelength and directly affects the temporal profile of the quantum states. For standard single-mode fibers at 1550~nm, the dispersion coefficient is approximately 20~ps/(nm$\cdot$km) \cite{djordjevic2025quantum}. Although dispersion-shifted fibers can reduce this effect, they may introduce nonlinear phenomena such as four-wave mixing when multiple wavelength channels are employed \cite{redyuk2025compensation}. Alternatively, digital-domain post-processing techniques, as discussed in Sec.~\ref{sec:4_post-proc}, can partially compensate for dispersion.

\begin{figure*}[!h]
    \centering
    \includegraphics[width=0.8\textwidth]{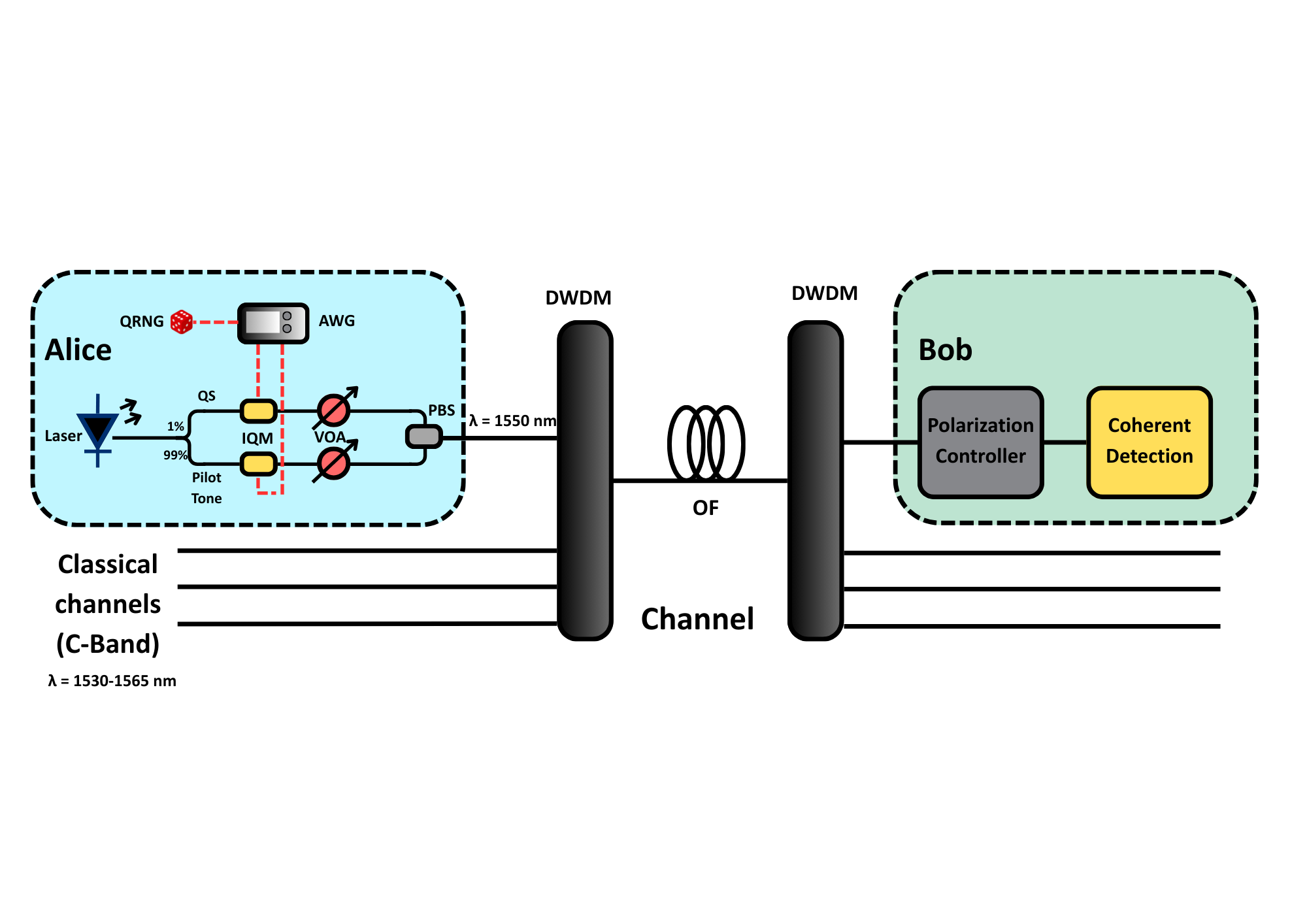}
    \caption{Coexistence of the quantum signal with classical C-Band ($\lambda = $ 1530-1565nm) channels. The signals are combined/separated with a dense wavelength-division multiplexer(DWDM). At the receiver, the quantum channel is demultiplexed from the classical traffic and coherently detected, with an active polarization control stage to compensate for polarization drifts induced by the optical fiber. }
    \label{fig:3_fiber_channel}
\end{figure*}

Another relevant impairment is polarization fluctuation. Optical fibers are sensitive to environmental perturbations, such as temperature changes and mechanical vibrations, which cause random polarization drifts. These fluctuations impact both the quantum signal and the pilot tone when they are transmitted simultaneously via polarization multiplexing, typically implemented with a polarization beam splitter (PBS). As illustrated in Fig.~\ref{fig:3_fiber_channel}, to transmit the quantum signal and the pilot tone simultaneously, we can use polarization multiplexing by combining both signals with a PBS. During propagation, the fiber induces polarization changes in both signals, requiring the use of a polarization control system. Most implementations rely on active polarization controllers \cite{chen2009stable}, which can introduce distortions into the optical setup. Alternatively, passive polarization controllers combined with Faraday mirrors have been employed to achieve long-term stabilization \cite{park20222}. On the other hand, it is possible to have a polarization-diverse detection. Although there will be leakage of the pilot tone into the quantum signal, it can be recovered using an equalizer in the post-processing step \cite{9927353}.

%Another issue as well is polarization. OFs are sensitive to temperature and vibration, and both can change the polarization of the quantum state. As depicted in Fig. \ref{fig:3_fiber_fso}, to transmit both the quantum signal and pilot tone simultaneously, they are combined into a polarization beam splitter (PBS). After the propagation, the OF will apply a change in polarization to both signals, making a polarization controlling system necessary. Most implementations utilize active polarization controllers, which introduce distortions into the optical systems; however, other references use passive polarization controllers in conjunction with Faraday Mirrors to maintain system stabilization \cite{nascimento2025passive}. 

Lastly, the coexistence with classical channels is another challenge \cite{hajomer2025coexistence}. With a larger quantity of classical channels, noises such as crosstalk and stimulated Raman scattering are more present. Crosstalk is the leakage of the classical channel into the quantum one. This leakage may originate from finite filter extinction ratios, non-ideal WDM components, or spectral broadening of classical signals due to modulation and fiber nonlinearities. As a consequence, crosstalk manifests as excess noise at the receiver, degrading the signal-to-noise ratio and potentially compromising the security of the quantum system. A possible solution for this problem is using spectral filters, such as a thinner Bragg grating, to diminish this leakage. Raman backscattering is a nonlinear effect of the fiber due to the interaction with two channels and the fiber structure, which generates phonons, therefore increasing the noise \cite{laudenbach2018continuous}. High-power classical channels can induce both spontaneous and stimulated Raman scattering, generating broadband noise photons that may spectrally overlap with the quantum channel. Raman noise is particularly detrimental due to its wide spectral distribution and its dependence on the relative wavelength spacing and launch power of the classical channels. While spectral filtering can partially mitigate this effect, a complementary approach is to use time multiplexing between classical and quantum channels, thereby preventing interaction between them. However, it will limit the rate of both channels.

A possible approach for QKD is to shift the quantum signal to the O-band, around a wavelength of 1310~nm \cite{gavignet2023co,wang2017long}. Operating in this band significantly reduces chromatic dispersion, crosstalk, and Raman scattering. However, fiber attenuation is higher in this wavelength range, typically around 0.3~dB/km \cite{wang2017long}. Consequently, the optimal operating regime depends on the trade-off between these effects and the achievable key rate.

%However, for longer distances, the exponential loss of optical fibers could be a problem. For this scenario, CV-QKD in free space has been proposed/implemented in the literature as a possible solution \cite{ghalaii2023continuous,li2020free,zhang2024continuous,zheng2025free}. FSO systems have a coupling loss that is higher than that of optical fibers. For short to metropolitan distances, optical fibers have better transmittance, but for longer distances and remote areas, FSO shows better results \cite{bakker2024best}. The main issue for FSO systems is the adaptive optics that are necessary to, in theory, recover the quantum signal that suffers from diffraction and the channel's turbulence, hence making it more complex to implement \cite{ghalaii2023continuous}. For optical fibers, the coexistence with classical channels is another problem \cite{hajomer2025coexistence}. With a larger quantity of classical channels, noises such as crosstalk and stimulated Raman scattering are more present. One solution adopted for QKD in general is transferring the quantum signal to the O-Band, i.e., around 1310 nm wavelength \cite{gavignet2023co,wang2017long}. This process makes the previously cited noises almost zero, but the fiber attenuation is higher for this band (typically around 0.3 dB/km) \cite{wang2017long}. Therefore, a careful analysis of which scenario one encounters should be done before implementing a CV-QKD protocol. Fig. \ref{fig:3_fiber_fso} explains the implementations of both channels.

\subsection{Receiver}

\begin{figure*}[!h]
    \centering
    \includegraphics[width=0.6\textwidth]{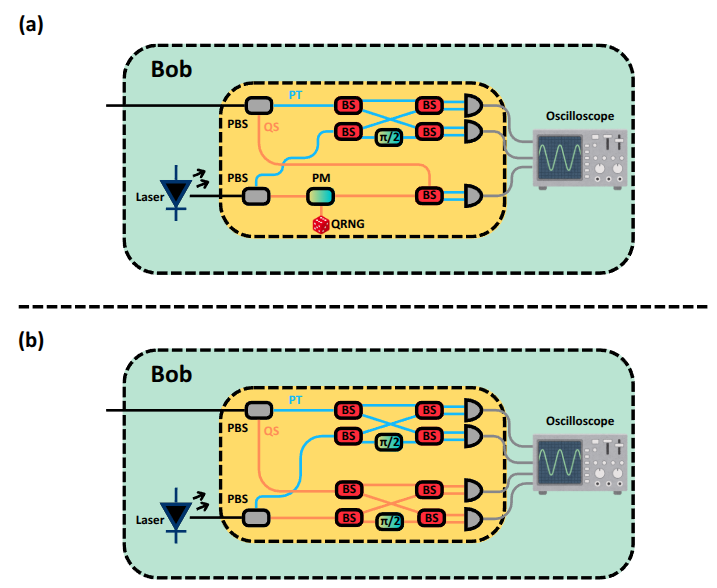}
    \caption{Detection setups for CV-QKD. In both figures, the blue line represents the pilot tone, while the orange line corresponds to the quantum signal. The gray boxes indicate the balanced detectors. (a) Homodyne detection. The quantum signal is separated from the pilot tone. It is directed into a beamsplitter together with the local-local oscillator (LLO), whose phase is randomized by a phase modulator (PM) driven by a quantum random number generator (QRNG). The pilot tone, on the other hand, passes through an interferometer with the LLO, where one branch introduces a phase shift of $\pi/2$, allowing full reconstruction of both quadratures of the pilot tone. (b) Double homodyne/heterodyne. In this configuration, both the quantum signal and the pilot tone enter an interferometer with the LLO, where one branch imposes a phase difference of $\pi/2$, enabling the full measurement of both quadratures for each signal. }
    \label{fig:3_det_system}
\end{figure*}

To measure information encoded in the optical quadratures, coherent detection techniques can be employed \cite{zhang2024continuous,laudenbach2018continuous}. In this context, two variants of the protocol can be implemented, as illustrated in Fig.~\ref{fig:3_det_system}.

The first variant relies on homodyne detection, in which Bob randomly selects one quadrature to measure from the quantum state transmitted by Alice. This measurement is performed using a homodyne detector combined with a phase modulator driven by a random number generator, which determines the quadrature to be analyzed \cite{zhang2024continuous}.

The second variant employs double homodyne detection, where Bob deterministically measures both quadratures of the incoming state. Although this approach enables simultaneous access to both quadratures, it incurs a penalty of 3~dB, corresponding to half of the original optical power, due to the use of an additional beam splitter to divide the signal \cite{laudenbach2018continuous}.

For both detection schemes, a balanced detector is required to ensure common-mode noise rejection and to isolate the quantum signal of interest \cite{laudenbach2018continuous,weedbrook2012gaussian}. After optical detection, the resulting electrical signal is forwarded to the post-processing stage.

Regardless of the detection method, detection noise will be present, and, according to Laudenbach et. al. \cite{laudenbach2018continuous}, it is the noise that mainly impacts CV-QKD systems. A balanced photodetector used for homodyne detection exhibits electronic noise from several internal components, including the photodiodes, transimpedance amplifier, and ADC. Each of these contributes with its own noise spectral density. Under this assumption, the total noise variance scales linearly with the detector bandwidth. Therefore, the increase in noise with bandwidth is not due to any specific detector parameter, but rather to the fact that a wider bandwidth integrates over a larger portion of the electronic noise spectrum. Therefore, if one chooses a detector with high electronic noise or insufficient bandwidth, it will significantly impact the secret key rate of the protocol.

Some studies demonstrate that we can treat detector noise as trusted, in which a portion of the observed noise is attributed to the detector itself rather than to a potential eavesdropper who cannot manipulate it. Incorporating this noise model enables more possibilities for analysis of the secure key rate in CV-QKD systems, accounting for imperfections inherent to the measurement devices. Despite its relevance, the detailed treatment of trusted detector noise is beyond the scope of this work. However, it has been reported in the following \cite{PhysRevResearch.3.043014,denys2021}.
%\subsection{Physical Implementation Challenges}\label{sec:Challenges}

\section{Post-processing Pipeline}
\label{sec:4_post-proc}
%\textbf{Micael, Davi Juvêncio e Nelson}

Post-processing in CV-QKD systems is an indispensable stage for transforming noisy and correlated measurements into secure secret keys. While the quantum channel is responsible for transmitting the modulated optical states, security against an adversary is effectively consolidated in the classical domain through posterior processing. During this stage, Alice and Bob apply protocols to reconcile the information with errors introduced by quantum noise, fiber attenuation, and excess noise, while simultaneously estimating and reducing the information potentially acquired by an eavesdropper. The efficiency of post-processing has a direct impact on both the final key generation rate and the maximum achievable distance, making it decisive for the feasibility of practical CV-QKD systems \cite{zhang2024continuous}.  

In general, the post-processing pipeline consists of a structured set of classical algorithms that follow the quantum transmission phase, typically including sifting, parameter estimation, information reconciliation, and privacy amplification \cite{zhang2024continuous}. Each stage plays a specific role in ensuring Alice and Bob derive identical and secret shared keys. However, sifting is not required when both quadratures are measured simultaneously using a no-switching protocol with dual quadrature homodyne or heterodyne detection, as all measurement results are retained for key generation \cite{Weedbrook2004WithoutSwitching,laudenbach2018continuous}.

In this context, a central bridge between quantum optical physics and post-processing is the digital signal processing (DSP) stage, which prepares the signal at the transmitter and processes the raw measurements at the receiver. This stage directly determines the quality of the correlations shared by Alice and Bob. While in the CV-QKD literature the post-processing part of the protocol is commonly related to the classical operations applied to the raw key to distill identical secret sequences, the DSP stage is responsible for recovering the measured signal by mitigating distortions caused by the channel and optical/electronic imperfections, thereby reducing excess noise. That is, DSP generates the raw data for key distillation. At the transmitter side, DSP encodes information into the signal by mapping bits into quadratures, organizes frames, inserts pilots, and applies pulse shaping (e.g., root-raised cosine filtering). At the receiver side, it separates pilot and quantum signals, performs digital down-conversion, clock recovery, static equalization, IQ imbalance correction, S21 response compensation, and dynamic equalization (e.g., phase tracking and polarization rotation via MIMO (multiple input multiple output)), as well as matched filtering and frame synchronization. Its control stack also performs shot-noise unit calibration. The outcome is a set of normalized and time-aligned samples with minimized excess noise and reduced bias for parameter estimation \cite{galvao2025nnexcessnoise}. In CV-QKD, this DSP chain — largely inherited from coherent communications — not only enables accurate LO stabilization and phase estimation, but also improves reconciliation efficiency $\beta$ and secure transmission distance, as it produces a more accurate covariance matrix by mitigating noise from system impairment. Recent works further highlight that even linear DSP routines must be carefully calibrated, particularly regarding consistent shot-noise estimation, to ensure overall security of the protocol ~\cite{10813518}.

\begin{figure*}[!h]
    \centering
    \includegraphics[width=1\textwidth]{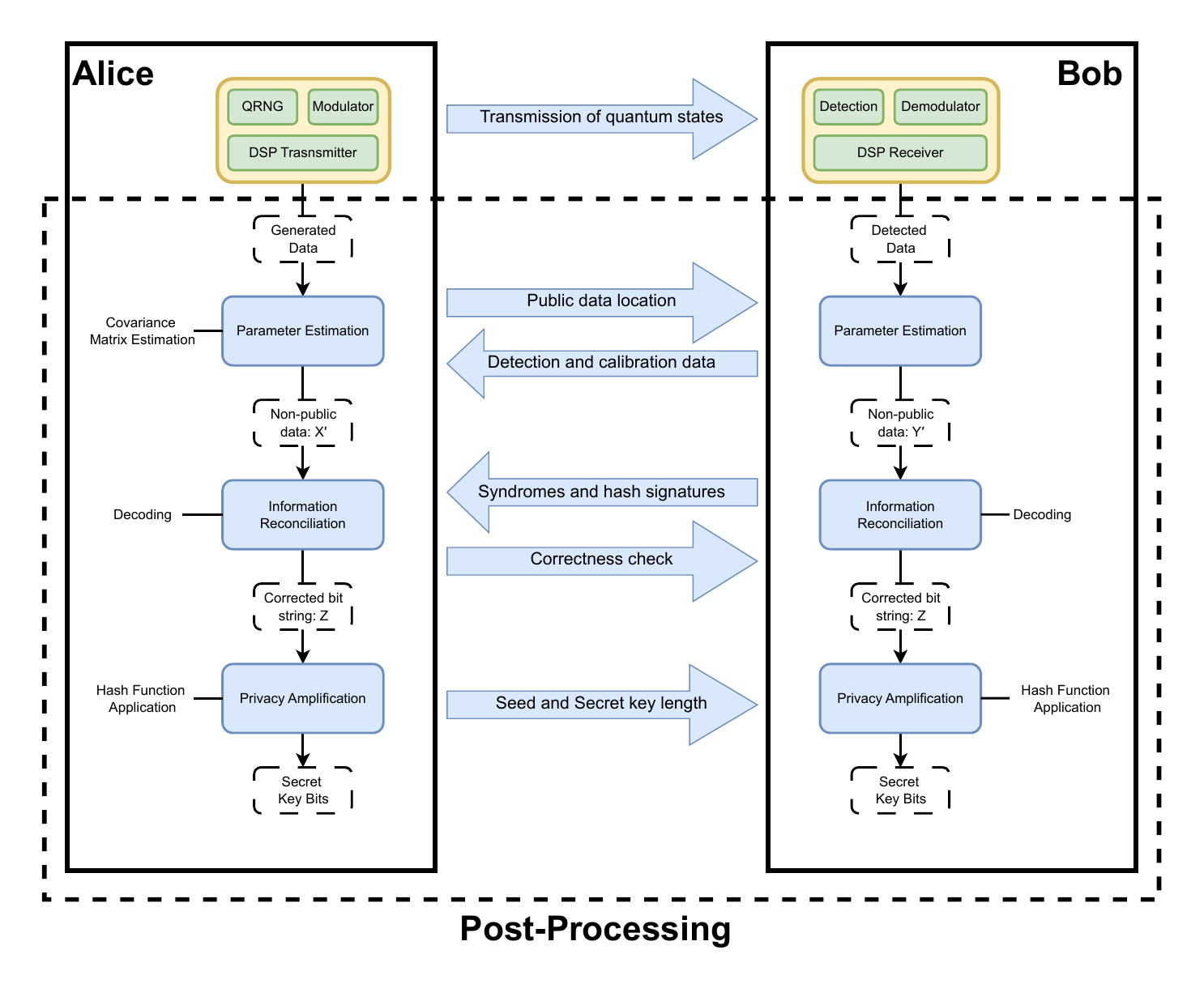}
    \caption{End-to-end pipeline of a CV-QKD system, illustrating the generation, transmission, detection, and post-processing of quantum states for secret key extraction. The upper part depicts the quantum layer, from state preparation at Alice to quadrature measurement at Bob, while the lower part shows the classical and authenticated post-processing stages, including parameter estimation, information reconciliation, and privacy amplification. Public communications and private data are explicitly distinguished. Adapted from \cite{zhang2024continuous}.}
    \label{fig:pipeline}
\end{figure*}

Fig.~\ref{fig:pipeline} illustrates the complete end-to-end flow. For clarity, the terms highlighted in bold correspond directly to the functional blocks explicitly labeled in the figure. On Alice’s side, the \textbf{QRNG} provides the raw key data, which are processed by the \textbf{DSP Transmitter}. This module maps the data onto optical quadratures by generating the baseband signal, inserting pilot symbols, and applying pulse shaping. 
The resulting digital samples are converted into electrical signals by the DAC and subsequently applied to the \textbf{Modulator}, which imprints them onto a continuous-wave laser. 
After appropriate attenuation, coherent states are generated and propagate through the quantum channel, as indicated by the transmission of quantum states region.

On Bob’s side, the incoming optical states are measured by the \textbf{Detection} stage, operating in either homodyne or heterodyne configuration. 
The measurement outcomes are then processed by the \textbf{DSP Receiver}, which performs digital demodulation tasks including down-conversion, clock recovery, equalization, and shot-noise-unit normalization, yielding the digital \textbf{Detected Data}.

Subsequently, the \textbf{Post-Processing} stage takes place and is entirely classical and authenticated. 
Alice and Bob first perform \textbf{Parameter Estimation} using publicly disclosed random subsets of their data ($X'$ and $Y'$) in order to reconstruct the covariance matrix and assess the viability of secret key extraction. 
They then execute \textbf{Information Reconciliation} to correct discrepancies between their strings and obtain the aligned bit sequence $Z$. 
Finally, after publicly communicating the \textbf{Seed and Secret key length}, both parties apply \textbf{Privacy Amplification}, based on Toeplitz hash functions, to generate identical \textbf{Secret Key Bits}.

The figure further highlights the distinction between public communications, such as announcements, syndromes, and correctness checks, and private data, while clearly indicating that the transmission of quantum states is restricted to the segment between modulation and detection.

\subsection{Digital Signal Processing for CV-QKD}
Digital Signal Processing (DSP) emerges as a pivotal technology in the paradigm shift from classical optical communications to quantum-secure systems, most notably in CV-QKD. At its core, DSP acts as the fundamental bridge between the quantum domain of information and the subsequent stages of classical post-processing, orchestrating the transformation of discrete sequences of quadrature values into a modulated optical signal at Alice’s station and their subsequent reconstruction into digital form at Bob’s station~\cite{schiavon2023high}. This capability is crucial not only for enabling data synchronization and mitigating signal impairments---such as chromatic dispersion, polarization-mode dispersion, and phase noise---in coherent optical communication systems, but also for enhancing the correlation between transmitted and received quantum states, thereby ensuring accurate key generation and the overall robustness of the CV-QKD system~\cite{10813518}. The adoption of DSP techniques originating from classical optical communications confers superior performance to CV-QKD in practical scenarios, where environmental factors and imperfections in optical components can degrade the quality of the quantum signal~\cite{10813518}. Moreover, DSP is indispensable for the implementation of a locally generated local oscillator (LO), enabling correction of frequency and phase offsets via frequency-multiplexed pilot tones transmitted by Alice~\cite{schiavon2023high}.

\begin{figure*}[!h]
    \centering
    \includegraphics[width=\textwidth]{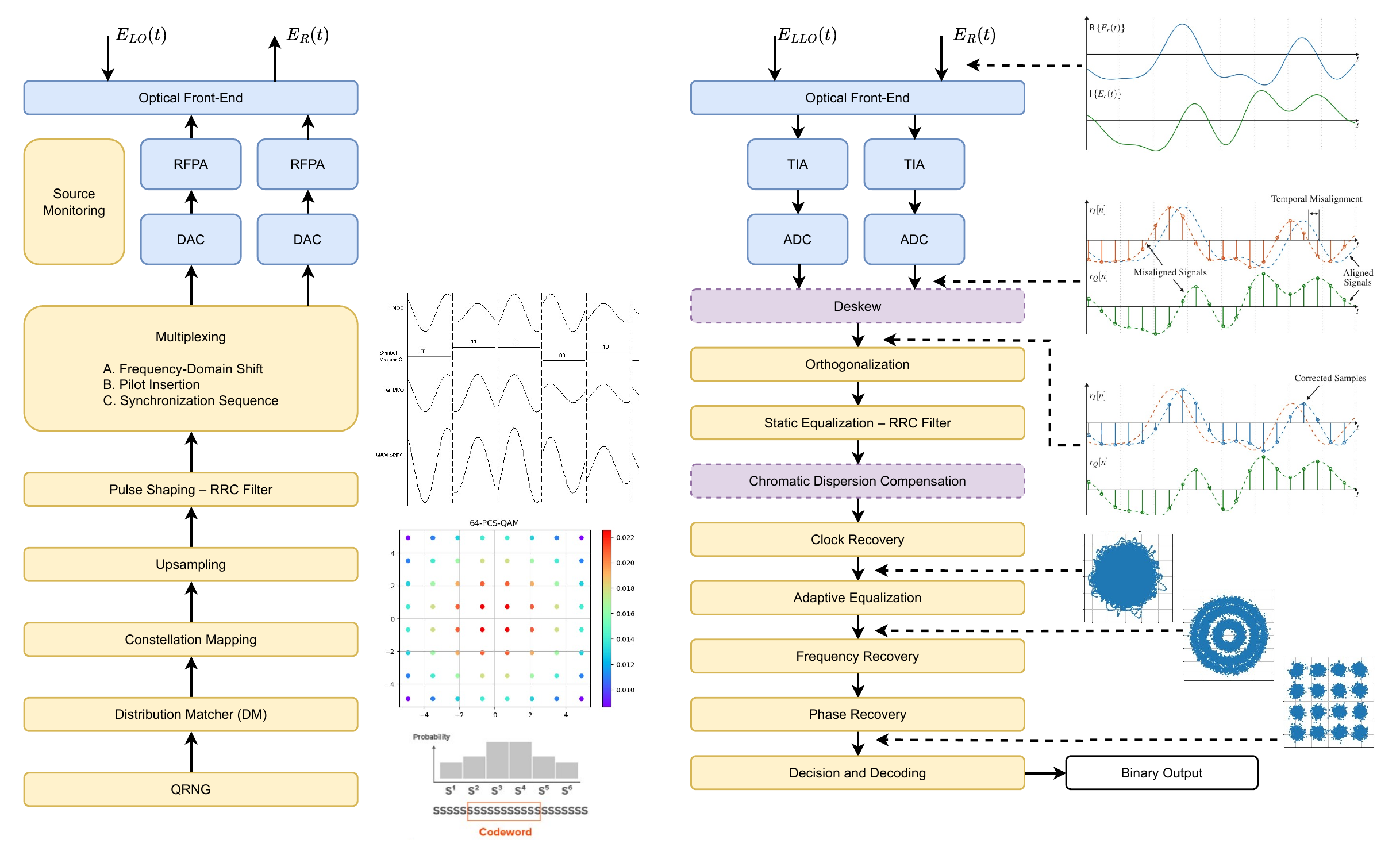}
    \caption{DSP pipeline of a CV-QKD system, from quantum state transmission to recovered raw data keys. Adapted from \cite{de2021digital}.}
    \label{fig:dsp_pipeline}
\end{figure*}

The DSP pipeline in CV-QKD, shown in Fig.~\ref{fig:dsp_pipeline}, represents the complete digital chain from quantum state preparation to key extraction. At the transmitter, the process begins with a QRNG, followed by distribution matching and constellation mapping, which determine the statistical properties of the encoded quadratures. The mapped symbols are pulse-shaped, typically with a root-raised cosine (RRC) filter, and digitally up-converted for optical modulation. Pilot tones and synchronization sequences are inserted to support frequency, phase, and timing recovery. At the receiver, the pipeline continues with high-speed analog-to-digital conversion (ADC), deskew, and static equalization. Chromatic dispersion compensation and adaptive equalization are then applied, followed by carrier frequency and phase recovery. Finally, decision and decoding modules convert the reconstructed quadratures into binary outputs that form the raw key material. This sequence highlights the role of DSP as the central element enabling reliable signal generation, synchronization, and impairment compensation in CV-QKD systems.

On the transmitter side, Alice’s DSP executes a series of critical functions to prepare the quantum signal for transmission. Random symbols are first generated from a prescribed distribution; for instance, in the Gaussian Modulated Coherent State (GMCS) protocol, the mean quadrature values $\langle I \rangle$ and $\langle Q \rangle$ are randomly drawn from a Gaussian distribution with zero mean and variance $V_A$~\cite{schiavon2023high}. Probabilistic Constellation Shaping (PCS) applied to Quadrature Amplitude Modulation (QAM) formats is also employed to optimize mutual information and approach the Shannon channel capacity~\cite{roumestan2024shaped}. Subsequently, pulse shaping is carried out, which is essential for defining the temporal mode of the signal. While an ideal time-limited pulse would require infinite bandwidth for distortion-free detection, engineering practice relies on spectrally limited pulse shapes such as the root-raised cosine (RRC) filter~\cite{roumestan2021high}. The RRC filter, with an adjustable roll-off factor, minimizes inter-symbol interference (ISI) and distortions inherent to finite-bandwidth detectors~\cite{schiavon2023high}. In experimental settings, the roll-off factor is also a key parameter influencing the system’s excess noise; therefore, it is typically optimized to values such as 0.4 or 0.2, depending on bandwidth constraints and overall performance requirements~\cite{hajomer2024continuous,roumestan2021high}. After pulse shaping, the signal undergoes digital frequency up-conversion primarily to mitigate low-frequency electronic noise, which would otherwise degrade the signal-to-noise ratio in the quadrature measurements. Although up-conversion also enables RF heterodyne detection, such detection does not strictly require this step, as it can alternatively be achieved by employing slightly detuned local oscillators at Alice’s and Bob’s sides. In RF heterodyne detection, the quantum signal is mixed with a frequency-shifted local oscillator, allowing simultaneous retrieval of both quadratures with a single photodiode, as demonstrated in recent true-heterodyne receiver architectures for CV-QKD~\cite{roumestan2021high,9927353}. To better reflect this context, references describing standard RF heterodyne implementations are more appropriate than those reporting phase-diverse detection schemes. Furthermore, frequency-multiplexed pilot tones are inserted into the quantum signal to provide robust references for clock and phase correction between Alice and Bob. A Zadoff-Chu or Constant Amplitude Zero Autocorrelation (CAZAC) sequence is appended at the beginning of each frame to facilitate synchronization and recovery of the time-multiplexed pilot sequence~\cite{zhang2024continuous}. Strategies such as frequency and polarization multiplexing of the pilots with the quantum signal are employed to mitigate crosstalk and channel interference~\cite{milovanvcev2021high}.

On the receiver side, Bob’s DSP is more complex and indispensable for faithful reconstruction of the quantum signal. The received electrical signals are first digitized by an analog-to-digital converter (ADC) operating at sampling rates ranging from 5~GSa/s to 10~GSa/s~\cite{roumestan2021high,pan2023simple}. The quantum signals and frequency-multiplexed pilot tones are then digitally separated~\cite{zhang2024continuous}. The quantum signal is subsequently down-converted to baseband, aligned with the frequency shift applied at the transmitter~\cite{zhang2024continuous}. Digital clock recovery identifies optimal sampling instants, employing algorithms such as Gardner’s method~\cite{xu2023real}. Static equalization then compensates for fixed imperfections, including I/Q imbalance and S21 mismatch, while the pulse is recovered through a matched RRC filter~\cite{pan2024100,zhang2024continuous}. Dynamic equalization is further applied to correct for variable phase and polarization drifts, frequently using Multiple-Input Multiple-Output (MIMO) algorithms~\cite{qiao2024novel}. This includes compensation of frequency offset, filtering and equalization, as well as pilot-aided phase recovery~\cite{dos2024real}. Frame synchronization is ensured through training sequences such as Zadoff-Chu or CAZAC~\cite{zhang2024continuous}, while fine data and clock synchronization are achieved via DSP-based routines~\cite{shen2023experimental}.

Beyond its basic functions, DSP in CV-QKD incorporates advanced techniques to optimize performance. Chromatic dispersion (CD) compensation, typically carried out in the frequency domain, mitigates pulse broadening effects in long-haul fiber transmission~\cite{dos2024real}. Phase and LO correction are enhanced by employing multiple frequency-multiplexed pilot tones and advanced carrier frequency/phase estimation algorithms, including the $M^{\text{th}}$-power method for residual phase error removal~\cite{schiavon2023high,roumestan2024shaped,chin2022digital}. To address polarization drifts and suppress noise, Kalman filters have shown effectiveness by adapting to slow variations in the polarization state~\cite{shen2023experimental}. A critical step is Shot Noise Unit (SNU) normalization, which provides a consistent and physically meaningful reference for quadrature measurements between Alice and Bob, rather than ensuring faithful quantum representation, and enables proper calibration of linear DSP outputs~\cite{chen2023continuous}.

The impact of DSP on CV-QKD system performance is multifaceted and deeply intertwined with security metrics. DSP optimization directly improves the reconciliation efficiency ($\beta$), which is crucial for maximizing the secret key rate by enhancing the correlation between Alice’s and Bob’s data~\cite{10813518}. Excess noise ($\xi$) is a limiting factor for both the achievable transmission distance and the security threshold of the protocol. Its reduction does not increase the intrinsic security of CV-QKD but ensures that the noise level remains below the maximum tolerable bound required for secure key generation~\cite{10813518}. The fidelity of the covariance matrix, constructed from DSP-recovered data, is essential for accurately assessing protocol security, as it enables estimation of the information potentially accessible to an eavesdropper (Eve)~\cite{chen2023continuous}. Consequently, the overall security of the protocol is constrained by DSP’s ability to maintain stable and accurate quadrature statistics while suppressing noise contributions that would otherwise compromise parameter estimation~\cite{10813518}.

Future perspectives for DSP in CV-QKD are promising, driven by the pursuit of higher speed, longer reach, and deeper integration. The application of machine learning (ML) algorithms for nonlinear channel equalization is an active research frontier, with neural networks demonstrating the potential to surpass conventional DSP approaches~\cite{freire2022neural,alshaer2024enhancing}. ML-based algorithms may also be applied to key sifting and excess noise reduction, enhancing system robustness in noisy environments~\cite{jin2021key}, as well as to automatic modulation format recognition~\cite{anjos2023fpga}. Integration into Application-Specific Integrated Circuits (ASICs) and Field-Programmable Gate Arrays (FPGAs) is fundamental for achieving real-time processing, as these platforms provide the parallel computational capabilities required by the intensive DSP operations, leading to significant increases in secret key rates and transmission distances~\cite{iqbal2024sdn}. The evolution toward system-on-chip (SoC) platforms, such as RFSoC-based architectures that integrate processors, programmable logic, and converters on a single chip, marks a decisive step in miniaturization and cost optimization, facilitating large-scale adoption of CV-QKD technology~\cite{anjos2023fpga}. Achieving real-time operation --- from quantum state preparation to final key extraction --- constitutes the ultimate goal, making overall system speed a primary performance indicator for commercial deployments~\cite{iqbal2024sdn}. Thus, the continuous evolution of DSP is intrinsic to the advancement of CV-QKD, driving the field toward increasingly efficient, secure, and adaptable quantum communication networks.

\subsection{Parameter Estimation and Calibration}\label{sec:parameter_estimation}

A fundamental security assumption for QKD protocols is that the quantum channel is insecure, meaning that the eavesdropper controls it and all information lost to the environment ultimately goes to the eavesdropper. In a practical scenario, this means that Eve controls the channel parameters, making precise parameter estimation a pivotal step in a QKD protocol operating in a non-asymptotic regime. Thus, Alice and Bob must use their random variables, $X$ and $Y$, respectively, to compute the worst-case-scenario parameters and
% To ensure unconditional security in the QKD scenario \cite{RevModPhys.74.145, RevModPhys.94.025008}, Alice ($X$) and Bob ($Y$) must not trust the channel specifications to 
construct their covariance matrix in order to compute the amount of information leaked to an eavesdropper ($E$) using the Holevo bound $\chi(Y; E)$ \cite{zhang2024continuous, usenko2025continuousvariablequantumcommunication, Diamanti2015}. 

The channel parameters must be estimated using a portion of the signals, denoted as $L'$, where $L$ is the total number of quadrature measurements after sifting (as introduced in Section~\ref{sec:2_theory}). The remaining $l = L - L'$ samples are reserved for raw key generation. In principle, they need to estimate the transmittance $T$, the excess noise $\xi$, Alice's variance $V_A$, quantum efficiency of the detectors $\eta_{\mathrm{eff}}$ and the electronic noise $\nu_{el}$. However, the main problem of parameter estimation is reduced to estimate $T$ and $\xi$, since one can reasonably assume that the other parameters are relatively well known:  \cite{PhysRevA.81.062343}. In fact, $\eta_{\text{eff}}$ is calibrated beforehand from the receiver's known efficiency and losses, and $V_A$ is a fixed parameter set and verified locally by the transmitter. In contrast, the excess noise has the most significant impact on the secret-key rate \cite{doi:10.1126/sciadv.adi9474, Huang2016}. 

The Gaussian channel parameters can be estimated considering the linear model
\begin{equation}
    Y = tX + Z \,,
\end{equation}
where $t = \sqrt T$, $X$ is a random variable related to the transmitted signal, $Z$ is a random variable related to the noise channel, and $Y$ is a random variable related to the received signal \cite{laudenbach2018continuous}.

In general, these variables are normalized in a shot-noise unit\textcolor{red}{s} (SNU) for both, Alice and Bob, have a reference unit for security analysis \cite{zhang2024continuous}. In this way, the variance of the quantum fluctuation in phase space is usually defined as this reference unit, represented by the SNU. In practical implementations, its estimation is made by measuring the vacuum, which is recorded in the receiver setup as electronic signals \cite{wang}. During the detection, the electronic noise is accounted to the shot-noise, such that Bob needs to make a calibration.

\begin{figure}
    \centering
    \includegraphics[width=1\linewidth]{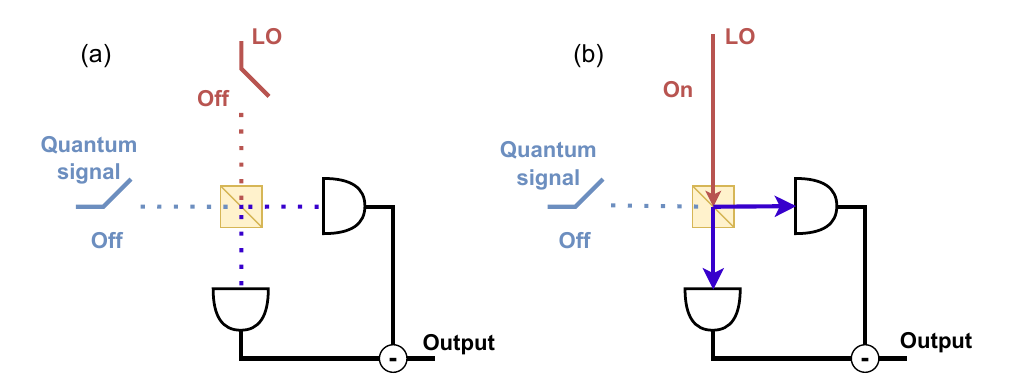}
    \caption{Calibrating a balanced homodyne detector. (a) Electronic noise level, measured by blocking both the signal and local oscillator (LO) inputs. (b) Total system noise, measured with the local oscillator (LO) beam applied.}
    \label{fig:Calibration}
\end{figure}

The calibration of shot noise can be performed using various methodologies  \cite{wang, Fossier_2009, Jouguet2013, Jouguet:12, Zhang_2019, PhysRevApplied.13.024058}. Here, we outline the primary strategies commonly employed in the literature:

\begin{enumerate}
    \item \textbf{Two-time with pre-calibration} \cite{Fossier_2009}: In this calibration, the electronic noise is measured (see Fig. \ref{fig:Calibration}). First, with the LO beam blocked at the optical input ports and the homodyne detector powered on, the measured variance corresponds to the raw electronic noise. Subsequently, the total system noise is measured with the LO illuminated. The shot-noise unit (SNU) is then obtained by calculating the difference between these two variance measurements.
    
    \item \textbf{Two-time in real-time} \cite{Jouguet2013, Jouguet:12}: in this calibration, the pre-calibration is also performed at first. This enables Bob to compute a linear relation between the optical power and the SNU. Then, a small portion of the LO is separated and constantly measured. Then, based on the computation made after pre-calibration, the SNU is adjusted in real-time. 
    
    \item \textbf{One-time} \cite{Zhang_2019, PhysRevApplied.13.024058}: In this method, the SNU is calibrated by measuring the total noise. Then, the signals are normalized considering this measure, such that the electronic noise can be represented using another vacuum state for security analisys. However, the performance of this protocol is slightly inferior to that of the traditional method because electronic noise is not estimated in this process. As a result, the electronic noise is considered untrusted, meaning the associated loss is also deemed untrusted.    
\end{enumerate}

Following calibration, the signals are normalized, and their quadratures are distributed as $X \sim \mathcal{N}(0, V_A)$, $Z  \sim \mathcal{N}(0, \sigma^2)$, and $Y \sim \mathcal{N}(0, t^2 V_A + \sigma^2)$, where $\sigma^2 = 1 + t^2\xi$ \cite{PhysRevA.81.062343}. In this expression, the unit variance ($1$) originates from the shot-noise normalization inherent to homodyne detection. For heterodyne detection, this value becomes $2$, as the signal is split on a balanced beam splitter to enable simultaneous measurement of both conjugate quadratures \cite{PhysRevResearch.3.043014}. The parameters $t$ and $\sigma^2$ are then estimated from the data, with the resulting estimators denoted as $\hat{t}$ and $\hat{\sigma}^2$.

% \sout{\mad{In the asymptotic scenario, the estimators are unbiased, such that $\mathbb{E}[\hat{t}] = t$ and $\mathbb{E}[\hat{\sigma}^2] = \sigma^2$ \cite{casella2024statistical}}}\MAD{A característica de um estimador ser enviesado (ou não) não depende da estatística ser tomada no cenário assintótico.}
An estimator is a function of samples of a random variable\footnote{More formally, the samples are a sequence of independent and identically distributed random variables.}, being a random variable itself, and the estimator $\hat\theta$ of a parameter $\theta$ is said to be unbiased if $\mathbb{E}[\hat{\theta}] = \theta$. However, in general, the variance of the estimator is a non-zero value which is inversely proportional to the number of samples provided to the estimator, meaning that the estimated value may differ from the true parameter value in a finite statistics setting.
% \sout{However, this assumption of perfect estimation is unattainable in practical, finite-size implementations}. 
In a CV-QKD scenario, one needs unbiased estimators for the channel parameters ($\mathbb{E}[\hat{t}] = t$ and $\mathbb{E}[\hat{\sigma}^2] = \sigma^2$), and provide that the achievable secret-key rate will not be overestimated \cite{PhysRevResearch.3.043014}. This can be done by computing the estimator's confidence intervals and using the worst case boundaries of such intervals as practical values to compute the secret key rate. 
% To guarantee this security, the parameter estimation must be analyzed using worst-case confidence intervals. 
In other words, one ensures that, with high probability, the parameter estimation satisfy $\hat{t}_{\text{min}} \leq t$ and $\hat{\sigma}^2_{\text{max}} \geq \sigma^2$, where $\hat{t}_{\text{min}}$ and $\hat{\sigma}^2_{\text{max}}$ are the upper and lower bounds of the confidence intervals for the transmittance and excess noise \cite{PhysRevA.81.062343}, respectively, given by

\begin{equation} \label{eq:t_min}
    t_{\min} \approx \hat t - z_{\epsilon_{PE}/2} \mathrm{SD}
\end{equation}
and 
\begin{equation} \label{eq:sigma_max}
  \sigma^2_{\max} \approx \hat \sigma^2 + z_{\epsilon_{PE}/2} \mathrm{SD},
\end{equation}
where $z_{\epsilon_{PE}/2} = \text{erf}^{-1}(1 - \epsilon_{PE}/2)$ and $\mathrm{erf}(x)$ is the error function \cite{casella2024statistical}. Here, $\mathrm{SD}$ is the standard deviation.

The primary challenge in practical implementations is to achieve highly precise parameter estimation using a limited subset of $m$ signals, as these are subsequently discarded because their information is announced over the authenticated classical channel \cite{Cai_2009}. Within the literature, Maximum Likelihood Estimation (MLE) is widely recognized as the standard method in the field \cite{PhysRevA.93.042343, PhysRevResearch.6.023321, PhysRevLett.125.010502}. This prominence is largely due to the first comprehensive security proof against Gaussian collective attacks in the finite-size regime being established using an MLE-based framework \cite{PhysRevA.81.062343}. Subsequently, alternative methods have been proposed to achieve improved performance, such as yielding tighter confidence intervals and thus higher secret key rates \cite{PhysRevA.93.042343, PhysRevA.97.022316, Luo_2022}. %More recently, the feasibility of employing neural networks for this task has been demonstrated, culminating in a security proof that provides composable security guarantees with a quantifiable failure probability $\epsilon_{\mathsf{PE}}$ in ref. \cite{galvao2025nnexcessnoise}.

\subsection{Information Reconciliation}\label{sec:information-reconciliation}

The objective of the Information Reconciliation (IR) process is to enable Alice and Bob to share a set of completely consistent key bits, even in the presence of inevitable discrepancies introduced during quantum transmission, whether caused by the physical characteristics of the channel or by eavesdropping attempts from third parties. To achieve this goal, classical error-correcting codes (EECs) are employed, allowing the identification and correction of discrepancies between the bit sequences initially obtained by each party. In this way, IR serves as a fundamental step to ensure the reliability of the final keys, reducing inconsistencies and guaranteeing the secure sharing of common keys \cite{zhang2024continuous}. 

However, the IR of CV-QKD is particularly challenging due to its operation in low SNR regimes over long distances. In addition to the incorporation of a digital signal processing (DSP) stage, the effective application of ECCs requires the development of novel design, optimized code designs that generate LDPC matrices suitable for real-time hardware implementation, with reduced latency and high throughput. Under low-SNR conditions, LDPC codes demand a high degree of redundancy to reliably correct errors and enhance the key parameters used to evaluate IR performance namely: reconciliation efficiency ($\beta$), frame error rate ($FER$) and throughput ($T$). The first is used to characterize the efficiency of the error correction and expressed by a factor $\beta \in [0,1]$, the second indicates the probability of IR failure (with lower values being preferable), and the third is directly associated with process performance, measuring the number of raw key bits processed per unit of time. In CV-QKD, given the transmission distance, higher reconciliation efficiency enables higher key rate\cite{yang2023information}.

Reconciliation can be implemented in two distinct forms with different performance characteristics: direct reconciliation (DR) or reverse reconciliation (RR). In DR, Alice sends the redundant information required for error correction to Bob, who uses it to correct the errors in his data  and obtain a bit string identical to Alice's. In RR, Bob's raw keys serve as the reference, and he sends the necessary error correction information to Alice enabling her to adjust her bit string to match his. RR can significantly extend the transmission distance and enable the generation of secure keys over longer ranges, making it the dominant approach.

The raw keys in CV-QKD are Gaussian variables, and several schemes have been proposed for their reconciliation. The most well-known are slice reconciliation, multidimensional reconciliation and signal reconciliation.

\begin{itemize}
    \item \textbf{Slice Reconciliation} - Suitable for relatively high SNR (greater than 0dB), typically in short transmission distances. In RR, Bob applies a quantization function $\mathcal{Q}: \mathbb{R} \rightarrow \{0, 1\}^m$ to transform each Gaussian variable $Y_i$ into an m-bit label $\{B_j(Y_i)\}$, $j = 1, ..., m$. Bob then employs a multi-level encoder, which encodes each bit level of the label independently as the syndromes of an ECC with coding rate $R_j$ $(1 \leq j \leq m$. To recover Bob's m-bit label ${B_j}$. Alice uses a multi-stage decoder, leveraging her own correlated varaible $X$ as side information. As a result, both parties ultimately share identical keys.
    \item \textbf{Multidimensional Reconciliation} - Suitable for low SNR (from -20dB to 0dB), typically occurring in long-distance CV-QKD. In this scenarios the raw keys of Alice and Bob are correlated Gaussian variables and the SNR will be very low for long transmission distances. In this case, the raw keys have a small absolute value and are distributed around 0. Thus, it is difficult to discriminate the sign and realize the encoding and decoding. The multidimensional reconciliation algorithm provides a powerful encoding scheme for low SNR scenario and thus effectively extend the key distribution distance. The channel between Alice and Bob is converted into a virtual biary input additive white Gaussian noise (AWGN) channel and therefore efficient binary codes can be employed .
    \item \textbf{Sign Reconciliation} - Direct encoding the continuous random variable to a key bit by using sign. The sign reconciliation has the feature of simplicity and low complexity, however the performance is low, due to some states are very close to each other, making discrimination difficult.
\end{itemize}

\begin{figure}[H]
    \centering
    \includegraphics[width=1.0\linewidth]{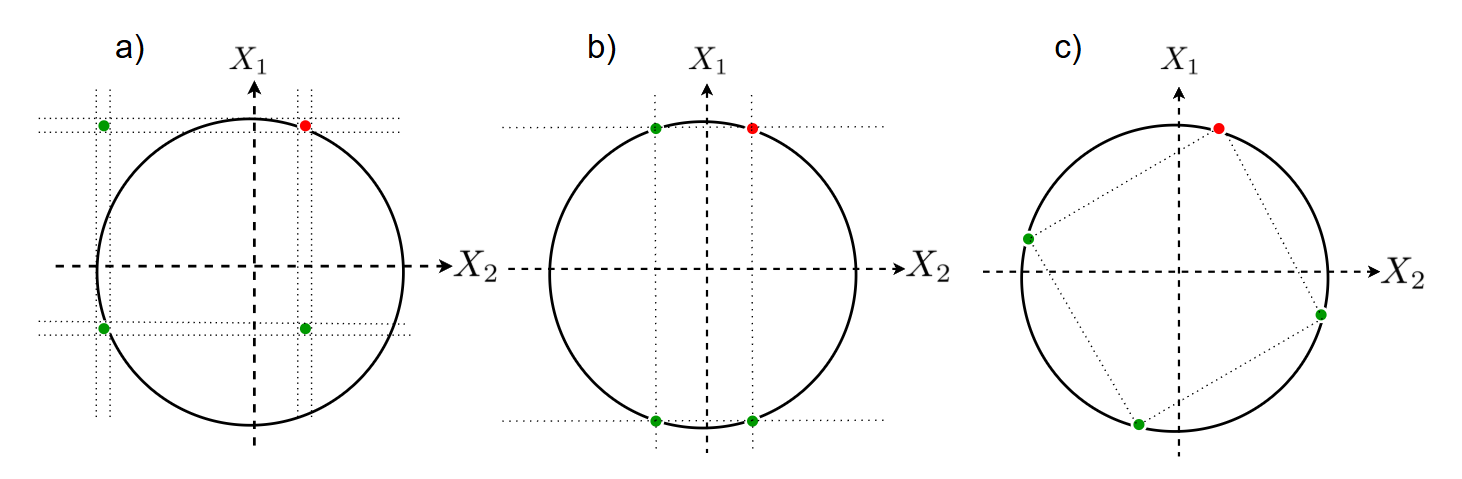}
    \caption{Figures (a), (b), and (c) illustrate the four possible states that Bob must discriminate after Alice sends side information through the authenticated classical channel. In (a), corresponding to slice reconciliation, the four states are well separated, but the Gaussian symmetry is broken. In (b), representing sign reconciliation, the problem’s symmetry is preserved, but some states are very close to each other, making discrimination difficult. In (c), corresponding to multidimensional reconciliation, a clever rotation is applied: the symmetry of the problem is preserved and the states become well to discriminate. }
    \label{fig: Fig7}
\end{figure}

The Figure \ref{fig: Fig7} shows the four possibles states that Bob needs to discriminate \cite{Leverrier_2008}. Consider that Alice and Bob share a set of correlated Gaussian variables $X$ and $Y$ and $x_1$ e $x_2$ belong to \textbf{x}. 

Multidimensional reconciliation has emerged as the most effective approach for low SNR regimes \cite{Leverrier_2008}, such as those encountered in long-distance CV-QKD. As previously discussed, this technique enables the use of efficient binary codes, such as LDPC, thereby reducing the error rate and enhancing the SKR.

Since CV-QKD systems typically operate over Gaussian channels with very low SNRs, approximately below 0 dB, the initial bit error rate is consequently very high. To mitigate this, ECCs with strong decoding performance are required to detect and correct errors introduced during transmission, thereby improving reliability. Most notably, LDPC codes, defined by sparse parity-check matrices $H$ of size $(n-k) \times n$, with code rate $R_{code} = \tfrac{k}{n}$. As mentioned earlier, in low-SNR regimes, very high redundancy is required to achieve reliable reconciliation, leading to large matrix dimensions (on the order of $10^6$) and significant computational burden during decoding. This complexity negatively impacts the performance of the decoding algorithm named as Belief Propagation (BP) and limits the real-time SKR, posing major challenge to the scalability of CV-QKD systems. Consequently, the design of LDPC code structures optimized for low-SNR and efficient hardware implementation is essential for advancing practical CV-QKD deployments \cite{yang2023information}.

LDPC codes are among the most promising strategies for application in the IR stage, as they enable performance close to the Shannon limit and are extensively employed in noisy communication systems. The parity-check matrix $H$ can be represented by a bipartite Tanner graph, denoted as $\mathcal{G}$, composed of two sets of nodes: the variable nodes $V_j$, associated with the columns of $H$, and the check nodes $C_i$, corresponding to its rows~\cite{moon2005error}. A connection between a variable node $j$ and a check node $i$ is established when $H(i,j) = 1$, indicating an edge between the nodes. The number of edges incident to a vertex of $\mathcal{G}$ is referred to as its degree, and the degree distribution is defined by two polynomials, $\lambda(x) = \sum_i \lambda_i x^i$ and $\rho(x) = \sum_i \rho_i x^i$, where $\lambda_i$ and $\rho_i$ denote the fraction of variable and check nodes of degree $i$, respectively~\cite{moon2005error}.

Alternative design approaches, for these codes are required in IR. In this context, Multi-Edge Type LDPC (MET-LDPC) proposed in \cite{Multi-edge-LDPCcodes}, generalize both irregular and irregular LDPC codes by introducing multiple edges types in the Tanner graph, allowing greater flexibility in defining the degree distribution of the nodes. This unified structure enables efficient modeling of codes with both uniform and non-uniform distributions, adapting them to varying channel conditions. A key advantage of MET-LDPC codes over conventional LDPC codes is their ability to achieve performance close to Shannon limit, particularly in low-rate regimes such as IR in CV-QKD systems, where they operate robustly even under extremely low SNR conditions. In addition, the LDPC Quasi-cyclic codes (QC-LDPC) represents a structure LDPC code class, in which the parity-check matrix $H$ is built from circulate blocks permutations of identity matrix. This structure allows a compact representation, facilitating efficient hardware implementations and significantly reduncing the computational cost of encoding and decoding, which is essential for CV-QKD systems with large block sizes \cite{milicevic2018quasi, li2020high}. These combined techniques can serve as a foundation for constructing codes applicable to various practical scenarios, including the optimization of IR in CV-QKD systems operating at very low SNRs, thereby extending the range and efficiency of QKD on long-distance optical communications.

\subsection{Privacy Amplification}
\label{PA}

Privacy amplification (PA) is integrated into the post-processing flow of CV-QKD systems, acting as the mechanism that converts the reconciled key into a final key uniformly random and entirely independent of any quantum system $E$ accessible to an eavesdropper \cite{zhang2024continuous,li2022quantum}. After the phases of \textit{sifting} (for discrete-variable protocols) or the conversion of continuous variables into binary bits (for CV-QKD), and the information reconciliation stage, Alice and Bob share a bit sequence that, although corrected, may still contain information potentially accessible to an eavesdropper (Eve) \cite{yan2022large}. The primary goal of PA is to compress the corrected sequence using a universal hash function into a significantly shorter sequence, thereby reducing Eve’s knowledge of the final key to an arbitrarily small level. This process is essential, since any residual base error rate is indistinguishable from eavesdropping, requiring that Eve’s information be either removed or rendered useless. The derivation of the final secure key length is extensively discussed in security analyses that take into account correctness and secrecy requirements \cite{li2022quantum, roumestan2021high, Furrer100502}.

The computational challenges in implementing PA are multifaceted and critical to the practical feasibility of high-speed CV-QKD systems. The complexity of universal hash algorithms is a limiting factor: while direct matrix multiplication has a computational complexity of \(O(N^2)\), techniques such as Toeplitz matrices, optimized by number-theoretic transforms (NTT) or fast Fourier transforms (FFT), can reduce this complexity to \(O(N \log N)\) \cite{li2022quantum}. However, these techniques produce uniform distributions only at the asymptotic regime, such that finite size effects of the order $O(n^{-1}\log_2(1/\epsilon_h))$ must be taken into account, where $\epsilon_h$ is the hashing parameter \cite{Konig2009}. To mitigate these effects, which inevitably reduce the secure key rate, the PA block size (input length, \(N\)) should be as large as possible, ideally \(10^{8}\) bits or more (e.g., \(10^{9},10^{10}\)), to effectively counter finite-size effects, improve the secure key rate, and approach asymptotic performance \cite{zhang2024continuous,yang2020high,yan2022large}. The need for processing rates on the order of gigabits per second (Gbps) is driven by the increasing repetition rates of CV-QKD systems and the adoption of wavelength-division multiplexing (WDM), which boosts raw key rates to unprecedented levels \cite{yan2022large}.

Given the large block sizes and stringent throughput requirements, real-time execution of privacy amplification demands dedicated hardware support. Conventional software-based approaches are unable to maintain the data rates required when operating on blocks approaching hundreds of millions of bits. Section~\ref{subsec:DHA} provides a detailed examination of the hardware trade-offs and architectural strategies necessary to support PA alongside information reconciliation and digital signal processing.

The impact of finite-size effects on the final secure key rate is a significant theoretical and practical concern \cite{yang2020high}. The security of CV-QKD is often analyzed in the asymptotic regime (infinite key length), but in practical scenarios, the key length is finite, leading to a reduction in the actual secure key rate. This reduction becomes more pronounced as part of the data must be allocated to parameter estimation, and statistical fluctuations grow with smaller block sizes. To mitigate these effects, it is essential to maximize the block length used for key generation, which in turn requires PA algorithms capable of processing extremely large data blocks. Simulations indicate that the improvement in secure key rate due to large-scale PA is particularly pronounced in CV-QKD systems, where finite-size effects are more critical than in DV-QKD \cite{yan2022large}.

% The impact of finite-size effects on the final secure key rate is a significant theoretical and practical concern. The security of CV-QKD is often analyzed in the asymptotic regime (infinite key length), but in practical scenarios the key length is finite, leading to a reduction in the actual secure key rate. The finite-size secure key rate \(K_{\text{finite}}\) is expressed as:
% \begin{equation}
% K_{\text{finite}} = \frac{n}{N} \big( \beta I_{AB} - \chi_{BE} - \Delta(n) \big)
% \label{eq:finite_key_rate}
% \end{equation}
% where \(n\) is the number of bits used for the key, \(N\) is the total sifted data, \(N-n\) is used for parameter estimation, \(\beta\) is the reconciliation efficiency, \(I_{AB}\) is the mutual information between Alice and Bob, \(\chi_{BE}\) is the Holevo bound, and \(\Delta(n)\) is the finite-size offset factor \cite{yang2020high}. To minimize the influence of \(\Delta(n)\), it is essential to maximize the length \(n\) of the processed key, which requires PA algorithms capable of handling extremely large data blocks. Simulations indicate that the improvement in secure key rate due to large-scale PA is particularly pronounced in CV-QKD systems, where finite-size effects are more critical than in DV-QKD \cite{yan2022large}.

\section{Hardware Development}
\label{sec:hardware_development}
The practical viability of CV-QKD systems, which can operate in real-time and over long distances, fundamentally depends on overcoming significant computational challenges. The DSP and post-processing stages, which ensure the key's security, require handling enormous volumes of data under low SNR conditions. To meet this demand, the development of dedicated hardware is necessary, as off-the-shelf solutions like CPUs and GPUs do not offer the required combination of performance, low power consumption, and latency to enable compact and integrable CV-QKD devices capable of generating keys in real-time on existing network infrastructures and being mass-produced.

\subsection{Digital Hardware Architecture}
\label{subsec:DHA}
The need for a dedicated, programmable, and flexible accelerator hardware architecture is a prerequisite for building CV-QKD systems that can evolve with research. The main objective is to enable real-time key generation and the efficient processing of large data volumes, overcoming the computational bottlenecks imposed by processing long transmission frames, a direct consequence of low SNR in the quantum regime and long-distance communications. A well-defined architecture is essential not only to achieve the necessary performance but also to ensure that the system can adapt to the continuous theoretical and algorithmic innovations that characterize the field of quantum communication.

In contrast to the concise reference in Section~\ref{PA}, which addressed privacy amplification in isolation, the assessment presented here encompasses the complete digital processing chain of CV-QKD systems. This broader scope is essential to evaluate the architectural requirements for real-time operation under practical deployment conditions.

The computational demands of a CV-QKD system are extraordinarily high. Both the DSP modules, responsible for signal encoding and recovery, and the post-processing stages — parameter estimation, IR, and PA — require massive processing power. To ensure security and efficiency in noisy channels, it is necessary to process data blocks that can exceed $10^9$ symbols \cite{Zhang_2019, Pirandola_2020, hajomer2024continuous}. Information reconciliation, for example, uses complex error correction codes (e.g., MET-LDPC) that operate on blocks of hundreds of millions of bits \cite{milicevic2018quasi, hajomer2024continuous}, while privacy amplification performs hash operations on matrices of equally massive dimensions, making real-time processing a major challenge \cite{zhang2024continuous}.

Off-the-shelf hardware solutions, such as CPUs and GPUs, are often insufficient to meet the stringent requirements of commercial CV-QKD systems, particularly for embedded or high-speed deployment \cite{Yang2020HighSpeed, Sun2024HighSpeedFPGA, ng2025gigabitqkd}. Although GPUs provide substantial parallelism and have been successfully employed in experimental demonstrations \cite{milicevic2018quasi,Weerasinghe2023}, their power consumption and latency characteristics limit their suitability for compact or energy-constrained applications. CPUs, in turn, lack the parallel processing capability needed to handle the data throughput in real-time. FPGAs emerge as a promising platform, combining low power consumption with the ability to efficiently manage both system control and computation, thereby facilitating integration with optical and electronic components \cite{li2022quantum,yan2022large}. Nonetheless, on-chip memory availability remains a critical constraint for FPGA-based implementations operating on very large code blocks. External memory interfaces, such as DDR, may be employed to extend capacity, but careful architectural planning is required to avoid throughput degradation. Furthermore, a hybrid architecture can be a good approach to address this need while accommodating the evolution of algorithms. In such an approach, computationally intensive and stable operations, including LDPC decoding and hashing, are implemented as dedicated hardware accelerators, while algorithmically evolving components, such as DSP functions and parameter estimation, execute on programmable processors. Application-Specific Instruction-set Processors (ASIPs), typically based on RISC-V extensions \cite{WatermanEECS2016_1, Dorflinger2021RISCV, ng2025gigabitqkd}, provide flexibility with a customized instruction set that accelerates key quantum-communication tasks without demanding hardware redesign.

Modularity is also an essential principle for an architectural concept. As CV-QKD theory continues to advance, DSP and post-processing algorithms are constantly being improved. A modular architecture allows individual system components to be updated or replaced independently. This partitioning strategy facilitates system maintenance and evolution, ensuring that the hardware platform can adapt to future generations of CV-QKD protocols.

The path for developing and implementing the digital hardware architecture should go through two distinct phases: prototyping and production. FPGAs are highly suitable for prototyping, validation, and integrated testing, owing to their reconfigurable nature and straightforward interface with optoelectronic subsystems \cite{Zhang2019PIC, anjos2023fpga}. After validating the concept on an FPGA, the final solution points towards the fabrication of an ASIC in the form of a SoC. This SoC would integrate the ASIPs and dedicated accelerators on a single chip, ensuring performance, low power consumption, and reduced unit cost necessary for mass production and the widespread adoption of CV-QKD technology.

\subsection{Software Framework}
\label{subsec:framework}
The development of such complex and specialized hardware requires, as an indispensable prerequisite, a robust and validated reference model. To this end, the creation of a comprehensive software framework is fundamental. This framework, named CV-QKD-ModSim, has been designed to model, simulate, and validate the entire digital information processing chain of a CV-QKD system, serving as a solid foundation and a blueprint for the hardware microarchitecture design. It allows exploring the design space and optimizing algorithms before committing resources to physical implementation.

The framework executes all digital processing steps, from the transmitter to the receiver. It simulates the DSP algorithms in Alice, such as pulse shaping and pilot tone multiplexing, and in Bob, such as clock recovery, equalization, and phase recovery. Furthermore, it models the signal propagation through the quantum channel, incorporating realistic effects like attenuation and excess noise from various sources, including laser phase noise and electronics imperfections. The framework implements the entire post-processing pipeline, including shot-noise unit (SNU) calibration and normalization, parameter estimation \cite{galvao2025nnexcessnoise}, information reconciliation (e.g., LDPC codes), and privacy amplification to ensure the security of the keys.

One of the most critical functions of CV-QKD-ModSim is to serve as a reference for transitioning algorithms from a floating-point model to a fixed-point model. The analysis of the required bit resolution for the numerical representation of each variable in the system is a determining step for hardware design. Insufficient precision can compromise security and performance, while excessive precision leads to a waste of logic resources, memory, and power in the FPGA or ASIC. The framework enables bit-true simulation, measures the impact of quantization, and determines the lowest binary word resolution that still guarantees secure key generation, thereby optimizing the efficiency of the final implementation.

The complexity of CV-QKD systems lies in the interdependence of multiple configuration parameters that need to be optimized simultaneously to maximize the secret key rate. The software framework provides a controlled environment to perform this optimization. Through computationally efficient simulations, it is possible to conduct a broad search in the parameter space, such as modulation variance, transmission frame size (e.g., blocks of $10^8$ up to $10^9$ symbols), and pilot tone strategies, to identify the optimal configuration for different channel conditions and distance scenarios. This fine-tuning process in software accelerates research and development, avoiding the cost and time associated with experimental tests implemented directly in hardware \cite{Pietri20242024, Kreinberg2020}.

The CV-QKD-ModSim software framework, being developed within the QuIIN research group, constitutes a core element of our hardware-oriented design methodology. It supports algorithm exploration and optimization, provides a validated digital reference model, and allows quantitative assessment of accuracy and resource requirements before deployment on FPGA or ASIC devices.

The current version is an internal prototype under continuous refinement, with its DSP and post-processing modules being validated through theoretical consistency checks and comparison with offline key-generation experiments performed on optical testbeds. As the framework reaches maturity, a public release under an open-source license is planned, along with a dedicated publication presenting its architecture, validation methodology, and benchmarking results. This roadmap ensures a reliable reference model for hardware design while maintaining alignment with practical system constraints.

%The software framework CV-QKD-ModSim being develeped by QuIIN researchers is not just a simulation tool but a central piece in the hardware development methodology for CV-QKD. It accelerates research, allows for the optimization of complex algorithms, de-risks the hardware design process by providing a validated reference model, and ultimately enables the creation of efficient and robust hardware accelerators capable of performing secure quantum key generation in real-time \cite{2, 12}.

% \section{Performance Analysis and Projections}

% \textbf{Braian e Nelson}

\section{Perspectives}
\label{sec:perspectives}

\subsection{QuIIN's Perspectives}
\label{sec:quiins_perspectives}
% \ABT{Colocar as promessas do QuIIN}

QuIIN's development plan for CV-QKD systems follows a staged approach toward practical deployment in real-world metropolitan networks. Our implementation will employ Gaussian modulation with heterodyne (dual-quadrature homodyne) detection. As a proof of concept, we will initially perform offline key extraction with GPU-accelerated processing, leveraging the frameworks discussed in Sec. \ref{subsec:framework}. The second phase involves prototyping a digital hardware FPGA/ASIC accelerator (see Sec. \ref{subsec:framework}) for real-time secret key generation. Following laboratory validation, we plan to conduct field tests on an operational metropolitan quantum network in partnership with the Rio Quantum Network (RQN), a quantum communication network under deployment in Rio de Janeiro~\cite{rederio, noticia_fapesp_RRQ_etc}. The RQN infrastructure connects major research institutions through both fiber-optic and free-space optical (FSO) channels, providing an ideal testbed for comprehensive evaluation of CV-QKD performance under realistic metropolitan conditions.

The validation phase on the RQN will evaluate our CV-QKD system under realistic metropolitan network conditions. The RQN's heterogeneous topology presents significant engineering challenges that must be addressed for practical implementation. Fiber links introduce chromatic dispersion and polarization mode dispersion that require compensation, while FSO segments face atmospheric attenuation, turbulence-induced beam wander, and pointing stability issues. Additionally, maintaining synchronization across the heterogeneous network poses a fundamental challenge. Characterizing system performance under these variable conditions is essential for validating the technology's readiness for broader deployment.

Following the RQN validation, our efforts will focus on optimizing FSO link performance to enable scaling beyond metropolitan networks. Key objectives include improving link margin, reducing beam divergence, and enhancing tracking accuracy through refined adaptive optics systems and robust modulation techniques that mitigate atmospheric effects. These improvements are necessary steps toward establishing reliable long-distance quantum channels that can support a future national quantum network infrastructure in Brazil, transitioning from isolated metropolitan deployments to integrated wide-area quantum communications.

Looking beyond the RQN validation phase, our research roadmap includes the investigation of CV-QKD applications in more challenging scenarios. Mobile platforms such as unmanned aerial vehicles (UAVs) represent an intermediate step toward space-based quantum communication, introducing engineering challenges including precision tracking of moving nodes, vibration compensation, and terminal miniaturization. Similarly, satellite-to-ground quantum links would enable global-scale secure communications but require significant advances in adaptive optics and link acquisition. These directions represent natural extensions of our FSO optimization efforts, though they remain subjects for future investigation as terrestrial CV-QKD technology reaches maturity.

\subsection{Scalability of quantum network}

The transition from isolated metropolitan quantum networks to a fully integrated national quantum infrastructure represents the next critical frontier for quantum communication in Brazil. While metropolitan networks in Rio de Janeiro, Recife, and São Carlos demonstrate QKD viability on a local scale, their interconnection across Brazil's vast and geographically diverse territory presents significant challenges due to distance limitations and exponential signal attenuation in optical fibers. Achieving national scalability requires moving beyond terrestrial fiber optics toward a multi-layer hybrid network architecture that integrates optical fibers, FSO links--including those enabled by drone-mounted or high-altitude platform station (HAPS) systems--and satellite-based quantum communication.

Long-haul FSO links represent the foundational step for national scalability, bridging intermediate distances between metropolitan areas or reaching regions where trenching fiber is impractical or prohibitively expensive. The partnership between QuIIN and the RQN, which plans to test a CV-QKD system over an 800-meter channel combining fiber and FSO, serves as a pilot for assessing CV-QKD performance under real-world FSO conditions characterized by atmospheric turbulence, scintillation, and variable attenuation \cite{mantey2025coexistence}.

For distances beyond a few hundred kilometers and coverage of remote regions, UAVs and HAPS offer dynamic and flexible solutions. Drone-based optical terminals can act as mobile trusted nodes, creating reconfigurable aerial mesh networks. The mobility of drones mitigates the challenge of maintaining line-of-sight for ground FSO by adapting to changing conditions. This layer can provide, for instance, rapid deployment for emergency communications and extend coverage to rural areas. 

Ultimately, satellite-based QKD provides the enabling layer for continental-scale coverage. As demonstrated by satellites such as Micius and Jinan-1, space-based links are the only viable technology for distributing keys over thousands of kilometers, connecting nodes across the entire national territory without requiring numerous intermediate trusted nodes \cite{chen2021integrated}. For Brazil, the choice between Low Earth Orbit (LEO) and Geostationary Orbit (GEO) satellites involves critical trade-offs. LEO satellites offer lower transmission losses but require complex tracking systems and handover protocols, which can introduce disruptions to the key stream \cite{alvaro2022caramuel, gao2024optimizing}. Conversely, GEO satellites provide stable, persistent links over vast geographic areas, which is highly advantageous for the prolonged interactive exchanges required by CV-QKD post-processing algorithms \cite{cheng2025feasibility}. However, this stability comes at the cost of significantly higher channel losses, demanding highly optimized systems with ultra-low noise detection \cite{orsucci2025assessment}. CV-QKD's coherent detection mechanism offers advantages in this space environment, as its narrow-band local oscillator inherently rejects broad-spectrum background radiation, enabling robust daytime operation and under strong stray light conditions \cite{derkach2020applicability}.

Realizing a scalable national quantum network requires a coordinated, multi-phase approach. This includes consolidating and standardizing existing metropolitan networks to ensure interoperability through common protocols and hardware interfaces. A strategic hybrid backbone must integrate fixed FSO links for inter-city connections alongside drone and HAPS-based networks for flexible coverage.

Concurrently, developing the space segment through national satellite capabilities or international partnerships is essential to secure space-based QKD resources. Thus, it requires determining the optimal orbital configuration, balancing LEO and GEO options, based on a thorough analysis of Brazil's specific needs for coverage, key rates, and latency requirements.

Finally, intelligent network management through sophisticated control plane software will dynamically route quantum key traffic across the hybrid ecosystem, continuously evaluating real-time channel conditions, weather data, and security requirements to select the most efficient and secure transmission path available among fiber, FSO, drone, and satellite links.

\section{Conclusion}\label{sec:Conclusion}

A century after quantum mechanics revolutionized our understanding of nature, we witness its transformation from theoretical framework to technological infrastructure. The second quantum revolution has matured from conceptual promise to engineering challenge. CV-QKD exemplifies this evolution, bridging quantum principles with classical telecommunication systems to deliver practical quantum security.

Realizing practical CV-QKD systems requires navigating substantial technical challenges across multiple layers. In this article, we have addressed these challenges comprehensively, from physical layer implementations to post-processing algorithms and hardware platforms. At the physical layer, the optical system must maintain phase stability while operating at extremely low SNR, where quantum effects dominate. Digital signal processing becomes critical for mitigating impairments in fiber links: chromatic dispersion and the ubiquitous phase noise that threatens coherent detection. Moving from offline GPU-accelerated post-processing to real-time FPGA-based key extraction demands not only computational power but algorithmic optimization that preserves security while maximizing throughput.

QuIIN's roadmap for developing Brazil's first CV-QKD system—from laboratory validation to field deployment on the RQN metropolitan infrastructure—directly confronts the gap between controlled environments and operational networks. Testing over real fiber infrastructure with unknown conditions will provide essential validation of CV-QKD's practical viability. Building a robust quantum technology ecosystem, however, requires more than infrastructure investment—it demands cultivating expertise that spans quantum fundamentals and engineering practice. Brazil's quantum future depends on professionals capable of navigating both domains: understanding the physics that governs secret key rates while engineering systems that operate reliably within practical and economic constraints.

As CV-QKD systems progress from research demonstrations to operational deployments, the challenges shift from ``can it work?" to ``can it scale?" The integration of quantum and classical systems, the coexistence with conventional data traffic, and the development of network protocols for quantum key management all require solutions grounded in both quantum theory and telecommunications engineering. As we mark this centennial of quantum mechanics, Brazil faces a defining choice in quantum technology development. The networks being deployed, the systems being engineered, and the professionals being trained today will determine whether Brazil contributes to or merely adopts quantum-secured communications infrastructure. 

\backmatter

\bmhead{Acknowledgements}

This work was partially funded by the projects: \textit{Non-conventional Receivers for CV-QKD, HW DSP: Development and Prototyping of Multicore SoC with Dedicated Accelerators and RISC-V DSP, Offline Demonstration for CV-QKD systems and LDPC Code Design for Information Reconciliation in CV-QKD Optimized for Hardware Implementation} supported by QuIIN – Quantum Industrial Innovation, the EMBRAPII CIMATEC Competence Center in Quantum Technologies, with financial resources from the PPI IoT/Industry 4.0 of the MCTI, grant number 053/2023, signed with EMBRAPII. MAD thanks financing from the European Union (HORIZON-MSCA-2023 Postdoctoral Fellowship, 101153602 - COCoVaQ).

\bmhead{Author Contributions Statement}
Introduction by Alexandre B. Tacla and Davi Juvêncio Gomes de Sousa; Overview on CV-QKD Systems by Micael Andrade Dias; Physical Layer by Christiano M. S. Nascimento; Post-processing Pipeline by Davi Juvêncio Gomes de Sousa, Nelson Alves Ferreira Neto, Lucas Q. Galvão, and Mauro Queiroz Nooblath Neto; Hardware Development by Davi Juvêncio Gomes de Sousa and Nelson Alves Ferreira Neto; and Perspectives by Braian Pinheiro da Silva and Cássio de Castro Silva. Conclusion by Alexandre B. Tacla. Conceptualization was carried out by Alexandre B. Tacla, Davi Juvêncio Gomes de Sousa and Valéria Loureiro da Silva. Alexandre B. Tacla and Davi Juvêncio Gomes de Sousa revised the manuscript.

\bmhead{Funding}
This work was partially funded by the projects, \textit{HW DSP: Development and Prototyping of Multicore SoC with Dedicated Accelerators and RISC-V DSP, Offline Demonstration for CV-QKD systems and LDPC Code Design for Information Reconciliation in CV-QKD Optimized for Hardware Implementation} supported by QuIIN – Quantum Industrial Innovation, the EMBRAPII CIMATEC Competence Center in Quantum Technologies, with financial resources from the PPI IoT/Industry 4.0 of the MCTI, grant number 053/2023, signed with EMBRAPII. MD thanks financing from the European Union (HORIZON-MSCA-2023 Postdoctoral Fellowship, 101153602 - COCoVaQ).

%\begin{appendices}

%\section{First appendix}\label{secA1}

%\end{appendices}

\bibliography{sn-bibliography}% common bib file

@misc{nguyen2025,
      title={Practical continuous-variable quantum key distribution with squeezed light}, 
      author={Huy Q. Nguyen and Ivan Derkach and Hou-Man Chin and Adnan A. E. Hajomer and Akash nag Oruganti and Radim Filip and Ulrik L. Andersen and Vladyslav C. Usenko and Tobias Gehring},
      year={2025},
      eprint={2506.19438},
      archivePrefix={arXiv},
      primaryClass={quant-ph},
      url={https://arxiv.org/abs/2506.19438}, 
}

@phdthesis{nguyen2025practical,
  author = {Nguyen, H. Q.},
  title = {Practical continuous variable quantum key distribution with squeezed light},
  school = {Technical University of Denmark},
  year = {2025},
  address = {Lyngby, Denmark},
  type = {Ph.D. dissertation}
}

@article{leverrier2015,
   title={Composable Security Proof for Continuous-Variable Quantum Key Distribution with Coherent States},
   volume={114},
   ISSN={1079-7114},
   url={http://dx.doi.org/10.1103/PhysRevLett.114.070501},
   DOI={10.1103/physrevlett.114.070501},
   number={7},
   journal={Physical Review Letters},
   publisher={American Physical Society (APS)},
   author={Leverrier, Anthony},
   year={2015},
   month=feb }

@article{jain2022,
   title={Practical continuous-variable quantum key distribution with composable security},
   volume={13},
   ISSN={2041-1723},
   url={http://dx.doi.org/10.1038/s41467-022-32161-y},
   DOI={10.1038/s41467-022-32161-y},
   number={1},
   journal={Nature Communications},
   publisher={Springer Science and Business Media LLC},
   author={Jain, Nitin and Chin, Hou-Man and Mani, Hossein and Lupo, Cosmo and Nikolic, Dino Solar and Kordts, Arne and Pirandola, Stefano and Pedersen, Thomas Brochmann and Kolb, Matthias and Ömer, Bernhard and Pacher, Christoph and Gehring, Tobias and Andersen, Ulrik L.},
   year={2022},
   month=aug }

@misc{anka2025,
      title={An introductory review of the theory of continuous-variable quantum key distribution: Fundamentals, protocols, and security}, 
      author={Maron F Anka and John A. Mora Rodríguez and Douglas F. Pinto and Lucas Q. Galvão and Micael A. Dias and Alexandre B. Tacla},
      year={2025},
      eprint={2512.01758},
      archivePrefix={arXiv},
      primaryClass={quant-ph},
      url={https://arxiv.org/abs/2512.01758}, 
}

@article{shannon1949,
  title = {Communication {{Theory}} of {{Secrecy Systems}}},
  author = {Shannon, Claude E.},
  year = {1949},
  journal = {Bell System Technical Journal},
  volume = {28},
  number = {4},
  pages = {656--715}
}

@article{nag2025continuous,
  title={Continuous-variable quantum key distribution with noisy squeezed states},
  author={nag Oruganti, Akash and Derkach, Ivan and Filip, Radim and Usenko, Vladyslav C},
  journal={Quantum Science and Technology},
  volume={10},
  number={2},
  pages={025023},
  year={2025},
  publisher={IOP Publishing}
}

@article{usenko2018unidimensional,
  title={Unidimensional continuous-variable quantum key distribution using squeezed states},
  author={Usenko, Vladyslav C},
  journal={Physical Review A},
  volume={98},
  number={3},
  pages={032321},
  year={2018},
  publisher={APS}
}

@book{djordjevic2019physical,
  author    = {Ivan B. Djordjevic},
  title     = {Physical-Layer Security and Quantum Key Distribution},
  publisher = {Springer},
  address   = {Cham},
  year      = {2019},
  edition   = {1},
  doi       = {10.1007/978-3-030-27565-5},
  isbn      = {978-3-030-27565-5},
  url       = {https://doi.org/10.1007/978-3-030-27565-5}
}

@article{sena2025,
    title={Um tutorial sobre Distribuição Quântica de Chaves: dos fundamentos às tecnologias modernas},
    author  = {Sena, Vitor L. and de~Melo, Fernando and Dias, Micael A. and Tacla, Alexandre B. and Chaves, Rafael},
    journal={Revista Brasileira de Ensino de Física},
    volume={vol. 47},
    pages={e20250373},
    year={2025}
}

@inproceedings{Bennett1984,
  title={Quantum cryptography: Public key distribution and coin tossing},
  author={Bennett, Charles H and Brassard, Gilles},
  booktitle={Proc. IEEE Int. Conf. on Computers, Systems and Signal Processing},
  pages={175–179},
  year={1984}
}

@article{Bedington2017,
   title={Progress in satellite quantum key distribution},
   volume={3},
   ISSN={2056-6387},
   url={http://dx.doi.org/10.1038/s41534-017-0031-5},
   DOI={10.1038/s41534-017-0031-5},
   number={1},
   journal={npj Quantum Information},
   publisher={Springer Science and Business Media LLC},
   author={Bedington, Robert and Arrazola, Juan Miguel and Ling, Alexander},
   year={2017},
   month=aug }

@article{li2025microsatellite,
  title={Microsatellite-based real-time quantum key distribution},
  author={Li, Yang and Cai, Wen-Qi and Ren, Ji-Gang and Wang, Chao-Ze and Yang, Meng and Zhang, Liang and Wu, Hui-Ying and Chang, Liang and Wu, Jin-Cai and Jin, Biao and others},
  journal={Nature},
  pages={47--54},
  year={2025},
  publisher={Nature Publishing Group UK London}
}

@article{Ren2017,
   title={Ground-to-satellite quantum teleportation},
   volume={549},
   ISSN={1476-4687},
   url={http://dx.doi.org/10.1038/nature23675},
   DOI={10.1038/nature23675},
   number={7670},
   journal={Nature},
   publisher={Springer Science and Business Media LLC},
   author={Ren, Ji-Gang and Xu, Ping and Yong, Hai-Lin and Zhang, Liang and Liao, Sheng-Kai and Yin, Juan and Liu, Wei-Yue and Cai, Wen-Qi and Yang, Meng and Li, Li and Yang, Kui-Xing and Han, Xuan and Yao, Yong-Qiang and Li, Ji and Wu, Hai-Yan and Wan, Song and Liu, Lei and Liu, Ding-Quan and Kuang, Yao-Wu and He, Zhi-Ping and Shang, Peng and Guo, Cheng and Zheng, Ru-Hua and Tian, Kai and Zhu, Zhen-Cai and Liu, Nai-Le and Lu, Chao-Yang and Shu, Rong and Chen, Yu-Ao and Peng, Cheng-Zhi and Wang, Jian-Yu and Pan, Jian-Wei},
   year={2017},
   month=aug, pages={70–73} }

@article{Yin2017,
author = {Juan Yin  and Yuan Cao  and Yu-Huai Li  and Sheng-Kai Liao  and Liang Zhang  and Ji-Gang Ren  and Wen-Qi Cai  and Wei-Yue Liu  and Bo Li  and Hui Dai  and Guang-Bing Li  and Qi-Ming Lu  and Yun-Hong Gong  and Yu Xu  and Shuang-Lin Li  and Feng-Zhi Li  and Ya-Yun Yin  and Zi-Qing Jiang  and Ming Li  and Jian-Jun Jia  and Ge Ren  and Dong He  and Yi-Lin Zhou  and Xiao-Xiang Zhang  and Na Wang  and Xiang Chang  and Zhen-Cai Zhu  and Nai-Le Liu  and Yu-Ao Chen  and Chao-Yang Lu  and Rong Shu  and Cheng-Zhi Peng  and Jian-Yu Wang  and Jian-Wei Pan },
title = {Satellite-based entanglement distribution over 1200 kilometers},
journal = {Science},
volume = {356},
number = {6343},
pages = {1140-1144},
year = {2017},
doi = {10.1126/science.aan3211},
url = {https://www.science.org/doi/abs/10.1126/science.aan3211}}

@article{Liao2017,
   title={Satellite-to-ground quantum key distribution},
   volume={549},
   ISSN={1476-4687},
   url={http://dx.doi.org/10.1038/nature23655},
   DOI={10.1038/nature23655},
   number={7670},
   journal={Nature},
   publisher={Springer Science and Business Media LLC},
   author={Liao, Sheng-Kai and Cai, Wen-Qi and Liu, Wei-Yue and Zhang, Liang and Li, Yang and Ren, Ji-Gang and Yin, Juan and Shen, Qi and Cao, Yuan and Li, Zheng-Ping and Li, Feng-Zhi and Chen, Xia-Wei and Sun, Li-Hua and Jia, Jian-Jun and Wu, Jin-Cai and Jiang, Xiao-Jun and Wang, Jian-Feng and Huang, Yong-Mei and Wang, Qiang and Zhou, Yi-Lin and Deng, Lei and Xi, Tao and Ma, Lu and Hu, Tai and Zhang, Qiang and Chen, Yu-Ao and Liu, Nai-Le and Wang, Xiang-Bin and Zhu, Zhen-Cai and Lu, Chao-Yang and Shu, Rong and Peng, Cheng-Zhi and Wang, Jian-Yu and Pan, Jian-Wei},
   year={2017},
   month=aug, pages={43–47} }

@article{chen2025implementation,
  title={Implementation of carrier-grade quantum communication networks over 10000 km},
  author={Chen, Hao-Ze and Li, Ming-Han and Wang, Yu Zhou and Zhao, Zhen-Geng and Ye, Cheng and Li, Fei Long and Chen, Zhu and Han, Sheng-Long and Tang, Bao and Miao, Ya Jun and others},
  journal={npj Quantum Information},
  volume={11},
  number={1},
  pages={137},
  year={2025},
  publisher={Nature Publishing Group UK London}
}

@article{liu2025road,
  title={The road to quantum internet: Progress in quantum network testbeds and major demonstrations},
  author={Liu, Jianqing and Le, Thinh and Ji, Tingxiang and Yu, Ruozhou and Farfurnik, Demitry and Byrd, Greg and Stancil, Daniel},
  journal={Progress in Quantum Electronics},
  volume={99},
  pages={100551},
  year={2025},
  publisher={Elsevier}
}

@article{wei2022towards,
  title={Towards real-world quantum networks: a review},
  author={Wei, Shi-Hai and Jing, Bo and Zhang, Xue-Ying and Liao, Jin-Yu and Yuan, Chen-Zhi and Fan, Bo-Yu and Lyu, Chen and Zhou, Dian-Li and Wang, You and Deng, Guang-Wei and others},
  journal={Laser \& Photonics Reviews},
  volume={16},
  number={3},
  pages={2100219},
  year={2022},
  publisher={Wiley Online Library}
}

@article{Cao2022,
  author={Cao, Yuan and Zhao, Yongli and Wang, Qin and Zhang, Jie and Ng, Soon Xin and Hanzo, Lajos},
  journal={IEEE Communications Surveys \& Tutorials}, 
  title={The Evolution of Quantum Key Distribution Networks: On the Road to the Qinternet}, 
  year={2022},
  volume={24},
  number={2},
  pages={839-894},
  doi={10.1109/COMST.2022.3144219}}

@article{park20222,
  title={2$\times$ N twin-field quantum key distribution network configuration based on polarization, wavelength, and time division multiplexing},
  author={Park, Chang Hoon and Woo, Min Ki and Park, Byung Kwon and Kim, Yong-Su and Baek, Hyeonjun and Lee, Seung-Woo and Lim, Hyang-Tag and Jeon, Seung-Woo and Jung, Hojoong and Kim, Sangin and others},
  journal={npj Quantum Information},
  volume={8},
  number={1},
  pages={48},
  year={2022},
  publisher={Nature Publishing Group UK London}
}

@article{redyuk2025compensation,
  title={Compensation of nonlinear signal distortions in optical fiber communication systems},
  author={Redyuk, Alexey and Sidelnikov, Oleg and Fedoruk, Mikhail},
  journal={Optics Communications},
  volume={578},
  pages={131418},
  year={2025},
  publisher={Elsevier}
}

@incollection{djordjevic2025quantum,
  author    = {Ivan B. Djordjevic},
  title     = {Physical-Layer Security, Quantum Key Distribution, and Post-Quantum Cryptography},
  publisher = {Springer Cham},
  year      = {2025},
  edition   = {2},
  isbn      = {978-3-031-88372-9},
  doi       = {10.1007/978-3-031-88372-9},
  url       = {https://doi.org/10.1007/978-3-031-88372-9}
}

@misc{wang2025,
      title={High-rate continuous-variable quantum key distribution over 100 km fiber with composable security}, 
      author={Heng Wang and Yang Li and Ting Ye and Li Ma and Yan Pan and Mingze Wu and Junhui Li and Yiming Bian and Yaodi Pi and Yun Shao and Jie Yang and Jinlu Liu and Ao Sun and Wei Huang and Stefano Pirandola and Yichen Zhang and Bingjie Xu},
      year={2025},
      eprint={2503.14843},
      archivePrefix={arXiv},
      primaryClass={quant-ph},
      url={https://arxiv.org/abs/2503.14843}, 
}

@article{silberhorn2002,
  title = {Continuous {{Variable Quantum Cryptography}}: {{Beating}} the 3 {{dB Loss Limit}}},
  shorttitle = {Continuous {{Variable Quantum Cryptography}}},
  author = {Silberhorn, {\relax Ch}. and Ralph, T. C. and L{\"u}tkenhaus, N. and Leuchs, G.},
  year = 2002,
  month = sep,
  journal = {Physical Review Letters},
  volume = {89},
  number = {16},
  pages = {167901},
  issn = {0031-9007, 1079-7114},
  doi = {10.1103/PhysRevLett.89.167901}
}

@article{hajomer2024,
author = {Adnan A. E. Hajomer  and Ivan Derkach  and Nitin Jain  and Hou-Man Chin  and Ulrik L. Andersen  and Tobias Gehring },
title = {Long-distance continuous-variable quantum key distribution over 100-km fiber with local local oscillator},
journal = {Science Advances},
volume = {10},
number = {1},
pages = {eadi9474},
year = {2024},
doi = {10.1126/sciadv.adi9474},
url = {https://www.science.org/doi/abs/10.1126/sciadv.adi9474},
}

@article{hajomer2024b,
  title = {Continuous-Variable Quantum Key Distribution at 10 {{GBaud}} Using an Integrated Photonic-Electronic Receiver},
  author = {Hajomer, Adnan A. E. and Bruynsteen, Cédric and Derkach, Ivan and Jain, Nitin and Bomhals, Axl and Bastiaens, Sarah and Andersen, Ulrik L. and Yin, Xin and Gehring, Tobias},
  date = {2024-09-20},
  journaltitle = {Optica},
  shortjournal = {Optica},
  volume = {11},
  number = {9},
  pages = {1197},
  issn = {2334-2536},
  doi = {10.1364/OPTICA.530080},
}

@misc{hajomer2025coexistence,
      title={Coexistence of continuous-variable quantum key distribution and classical data over 120-km fiber}, 
      author={Adnan A. E. Hajomer and Ivan Derkach and Vladyslav C. Usenko and Ulrik L. Andersen and Tobias Gehring},
      year={2025},
      eprint={2502.17388},
      archivePrefix={arXiv},
      primaryClass={quant-ph},
      url={https://arxiv.org/abs/2502.17388}, 
}

@misc{hajomer2025chipbased,
      title={Chip-Based 16 GBaud Continuous-Variable Quantum Key Distribution}, 
      author={Adnan A. E. Hajomer and Axl Bomhals and CÉdric Bruynsteen and Aboobackkar Sidhique and Ivan Derkach and Ulrik L. Andersen and Xin Yin and Tobias Gehring},
      year={2025},
      eprint={2504.09308},
      archivePrefix={arXiv},
      primaryClass={quant-ph},
      url={https://arxiv.org/abs/2504.09308}, 
}

@misc{noticia_fapesp_RRQ_etc,
  title = {Brazil’s first quantum cryptography network is expected to connect five research institutions},
  howpublished = {\url{https://revistapesquisa.fapesp.br/en/brazils-first-quantum-cryptography-network-is-expected-to-connect-five-research-institutions/}},
  note = {This story was published with the title “Qubits in Guanabara” in Revista Pesquisa FAPESP, issue 342 of august 2024.}
}

@inproceedings{rederio,
 author = {Guilherme Temporão and Fernando Melo and Antonio Khoury},
 title = { The Rio Quantum Network: a reconfigurable hybrid multi-user metropolitan quantum key distribution network},
 booktitle = {Anais do I Workshop de Redes Quânticas},
 location = {Niterói/RJ},
 year = {2024},
 issn = {0000-0000},
 pages = {19--24},
 publisher = {SBC},
 address = {Porto Alegre, RS, Brasil},
 doi = {10.5753/wqunets.2024.2872}
}

@article{deutsch20,
  title = {Harnessing the Power of the Second Quantum Revolution},
  author = {Deutsch, Ivan H.},
  journal = {PRX Quantum},
  volume = {1},
  issue = {2},
  pages = {020101},
  numpages = {13},
  year = {2020},
  month = {Nov},
  publisher = {American Physical Society},
  doi = {10.1103/PRXQuantum.1.020101},
  url = {https://link.aps.org/doi/10.1103/PRXQuantum.1.020101}
}

@article{zhang2024continuous,
  title={Continuous-variable quantum key distribution system: Past, present, and future},
  author={Zhang, Yichen and Bian, Yiming and Li, Zhengyu and Yu, Song and Guo, Hong},
  journal={Applied Physics Reviews},
  volume={11},
  number={1},
  year={2024},
  publisher={AIP Publishing}
}

@article{chen2009stable,
  title={Stable quantum key distribution with active polarization control based on time-division multiplexing},
  author={Chen, Jing and Wu, G and Xu, L and Gu, X and Wu, E and Zeng, H},
  journal={New Journal of Physics},
  volume={11},
  number={6},
  pages={065004},
  year={2009},
  publisher={IOP Publishing}
}

@article{assche2004,
  title = {Reconciliation of a Quantum-Distributed {{Gaussian}} Key},
  author = {Assche, Gilles Van and Cardinal, J. and Cerf, Nicolas J.},
  year = {2004},
  journal = {IEEE TIT},
  volume = {50},
  number = {2},
  pages = {394--400},
  issn = {0018-9448},
  doi = {10.1109/tit.2003.822618}
}

@article{grosshans2002,
  title = {Continuous {{Variable Quantum Cryptography Using Coherent States}}},
  author = {Grosshans, Fr{\'e}d{\'e}ric and Grangier, Philippe},
  year = {2002},
  month = jan,
  journal = {Physical Review Letters},
  volume = {88},
  number = {5},
  pages = {57902},
  publisher = {American Physical Society},
  doi = {10.1103/PhysRevLett.88.057902}
}

@article{manimozhi2025,
  title = {Post-Quantum {{AES}} Encryption Using {{ECC}} Points Derived from {{BB84}} Sifted Keys},
  author = {Manimozhi, M. and Mugelan, R. K.},
  year = {2025},
  month = dec,
  journal = {EPJ Quantum Technology},
  volume = {12},
  number = {1},
  pages = {109},
  issn = {2662-4400, 2196-0763},
  doi = {10.1140/epjqt/s40507-025-00411-9}
}

@article{zhang2022,
  title = {Quantum-Key-Expansion Protocol Based on Number-State-Entanglement-Preserving Tensor Network with Compression},
  author = {Zhang, Qiang and Lai, Hong and Pieprzyk, Josef},
  year = {2022},
  month = mar,
  journal = {Physical Review A},
  volume = {105},
  number = {3},
  pages = {032439},
  issn = {2469-9926, 2469-9934},
  doi = {10.1103/PhysRevA.105.032439}
}

@article{usenko2016,
  title = {Trusted {{Noise}} in {{Continuous-Variable Quantum Key Distribution}}: {{A Threat}} and a {{Defense}}},
  shorttitle = {Trusted {{Noise}} in {{Continuous-Variable Quantum Key Distribution}}},
  author = {Usenko, Vladyslav and Filip, Radim},
  year = {2016},
  month = jan,
  journal = {Entropy},
  volume = {18},
  number = {1},
  pages = {20},
  publisher = {MDPI AG},
  issn = {1099-4300},
  doi = {10.3390/e18010020},
  copyright = {https://creativecommons.org/licenses/by/4.0/}
}

@article{ghorai2019,
  title = {Asymptotic {{Security}} of {{Continuous-Variable Quantum Key Distribution}} with a {{Discrete Modulation}}},
  author = {Ghorai, Shouvik and Grangier, Philippe and Diamanti, Eleni and Leverrier, Anthony},
  year = {2019},
  month = jun,
  journal = {Physical Review X},
  volume = {9},
  number = {2},
  pages = {021059},
  issn = {2160-3308},
  doi = {10.1103/PhysRevX.9.021059}
}

@article{djordjevic2019,
  title = {Optimized-{{Eight-State CV-QKD Protocol Outperforming Gaussian Modulation Based Protocols}}},
  author = {Djordjevic, Ivan B.},
  year = {2019},
  month = aug,
  journal = {IEEE Photonics Journal},
  volume = {11},
  number = {4},
  pages = {1--10},
  issn = {1943-0655, 1943-0647},
  doi = {10.1109/JPHOT.2019.2921521}
}

@article{denys2021,
  title = {Explicit Asymptotic Secret Key Rate of Continuous-Variable Quantum Key Distribution with an Arbitrary Modulation},
  author = {Denys, Aur{\'e}lie and Brown, Peter and Leverrier, Anthony},
  year = {2021},
  month = sep,
  journal = {Quantum},
  volume = {5},
  pages = {540},
  issn = {2521-327X},
  doi = {10.22331/q-2021-09-13-540}
}

@article{soh2015,
  title = {Self-{{Referenced Continuous-Variable Quantum Key Distribution Protocol}}},
  author = {Soh, Daniel B. S. and Brif, Constantin and Coles, Patrick J. and L{\"u}tkenhaus, Norbert and Camacho, Ryan M. and Urayama, Junji and Sarovar, Mohan},
  year = {2015},
  journal = {Phys. Rev. X},
  volume = {5},
  number = {4},
  publisher = {American Physical Society (\{APS\})},
  doi = {10.1103/physrevx.5.041010}
}

@article{pereira2022,
  title = {Probabilistic Shaped 128-{{APSK CV-QKD}} Transmission System over Optical Fibres},
  author = {Pereira, Daniel and Almeida, Margarida and Fac{\~a}o, Margarida and Pinto, Armando N. and Silva, Nuno A.},
  year = {2022},
  month = aug,
  journal = {Optics Letters},
  volume = {47},
  number = {15},
  pages = {3948},
  issn = {0146-9592, 1539-4794},
  doi = {10.1364/OL.456333}
}

@article{bai2017,
  title = {High-Efficiency Reconciliation for Continuous Variable Quantum Key Distribution},
  author = {Bai, Zengliang and Yang, Shenshen and Li, Yongmin},
  year = {2017},
  month = mar,
  journal = {Japanese Journal of Applied Physics},
  volume = {56},
  number = {4},
  pages = {44401},
  publisher = {Japan Society of Applied Physics},
  doi = {10.7567/jjap.56.044401}
}

@article{laudenbach2018continuous,
  title={Continuous-variable quantum key distribution with Gaussian modulation—the theory of practical implementations},
  author={Laudenbach, Fabian and Pacher, Christoph and Fung, Chi-Hang Fred and Poppe, Andreas and Peev, Momtchil and Schrenk, Bernhard and Hentschel, Michael and Walther, Philip and H{\"u}bel, Hannes},
  journal={Advanced Quantum Technologies},
  volume={1},
  number={1},
  pages={1800011},
  year={2018},
  publisher={Wiley Online Library}
}

@article{wang2017long,
  title={Long-distance copropagation of quantum key distribution and terabit classical optical data channels},
  author={Wang, Liu-Jun and Zou, Kai-Heng and Sun, Wei and Mao, Yingqiu and Zhu, Yi-Xiao and Yin, Hua-Lei and Chen, Qing and Zhao, Yong and Zhang, Fan and Chen, Teng-Yun and others},
  journal={Physical Review A},
  volume={95},
  number={1},
  pages={012301},
  year={2017},
  publisher={APS}
}

@article{gavignet2023co,
  title={Co-propagation of QKD \& 6 Tb/s (60$\times$ 100G) DWDM channels with~ 17 dBm total WDM power in single and multi-span configurations},
  author={Gavignet, Paulette and Pincemin, Erwan and Herviou, Fabrice and Loussouarn, Yann and Mondain, Fran{\c{c}}ois and Grant, Andy and Johnson, Lee and Woodward, Robert Ian and Dynes, James Francis and Summers, Benedict and others},
  journal={Journal of Lightwave Technology},
  volume={42},
  number={4},
  pages={1321--1327},
  year={2023},
  publisher={IEEE}
}

@article{ghalaii2023continuous,
  title={Continuous-variable measurement-device-independent quantum key distribution in free-space channels},
  author={Ghalaii, Masoud and Pirandola, Stefano},
  journal={Physical Review A},
  volume={108},
  number={4},
  pages={042621},
  year={2023},
  publisher={APS}
}

@article{weedbrook2012gaussian,
  title={Gaussian quantum information},
  author={Weedbrook, Christian and Pirandola, Stefano and Garc{\'\i}a-Patr{\'o}n, Ra{\'u}l and Cerf, Nicolas J and Ralph, Timothy C and Shapiro, Jeffrey H and Lloyd, Seth},
  journal={Reviews of Modern Physics},
  volume={84},
  number={2},
  pages={621--669},
  year={2012},
  publisher={APS}
}

@article{qi2015generating,
  title={Generating the local oscillator “locally” in continuous-variable quantum key distribution based on coherent detection},
  author={Qi, Bing and Lougovski, Pavel and Pooser, Raphael and Grice, Warren and Bobrek, Miljko},
  journal={Physical Review X},
  volume={5},
  number={4},
  pages={041009},
  year={2015},
  publisher={APS}
}

@article{jouguet2013preventing,
  title={Preventing calibration attacks on the local oscillator in continuous-variable quantum key distribution},
  author={Jouguet, Paul and Kunz-Jacques, S{\'e}bastien and Diamanti, Eleni},
  journal={Physical Review A—Atomic, Molecular, and Optical Physics},
  volume={87},
  number={6},
  pages={062313},
  year={2013},
  publisher={APS}
}

@INPROCEEDINGS{10813518,
  author={da Silva, Valéria Loureiro and Dias, Micael Andrade and Neto, Nelson Alves Ferreira and Tacla, Alexandre B.},
  booktitle={2024 SBFoton International Optics and Photonics Conference (SBFoton IOPC)}, 
  title={From Coherent Communications to Quantum Security: Modern Techniques in CV-QKD}, 
  year={2024},
  volume={},
  number={},
  pages={1-5},
  doi={10.1109/SBFotonIOPC62248.2024.10813518}}

@article{Weedbrook2004WithoutSwitching,
  title = {Quantum Cryptography Without Switching},
  author = {Weedbrook, Christian and Lance, Andrew M. and Bowen, Warwick P. and Symul, Thomas and Ralph, Timothy C. and Lam, Ping Koy},
  journal = {Phys. Rev. Lett.},
  volume = {93},
  issue = {17},
  pages = {170504},
  numpages = {4},
  year = {2004},
  month = {Oct},
  publisher = {American Physical Society},
  doi = {10.1103/PhysRevLett.93.170504},
  url = {https://link.aps.org/doi/10.1103/PhysRevLett.93.170504}
}

@article{chen2023continuous,
  title={Continuous-mode quantum key distribution with digital signal processing},
  author={Chen, Ziyang and Wang, Xiangyu and Yu, Song and Li, Zhengyu and Guo, Hong},
  journal={npj Quantum Information},
  volume={9},
  number={1},
  pages={28},
  year={2023},
  publisher={Nature Publishing Group UK London}
}

@inproceedings{schiavon2023high,
  title={High-speed continuous-variable quantum key distribution with advanced digital signal processing},
  author={Schiavon, Matteo and Pi{\'e}tri, Yoann and Vidarte, Luis Trigo and Fruleux, Damien and Huguenot, Manon and Gouraud, Baptiste and Rhouni, Amine and Grangier, Philippe and Diamanti, Eleni},
  booktitle={2023 23rd International Conference on Transparent Optical Networks (ICTON)},
  pages={1--6},
  year={2023},
  organization={IEEE}
}

@article{roumestan2024shaped,
  title={Shaped constellation continuous variable quantum key distribution: Concepts, methods and experimental validation},
  author={Roumestan, Francois and Ghazisaeidi, Amirhossein and Renaudier, Jeremie and Vidarte, Luis Trigo and Leverrier, Anthony and Diamanti, Eleni and Grangier, Philippe},
  journal={Journal of Lightwave Technology},
  volume={42},
  number={15},
  pages={5182--5189},
  year={2024},
  publisher={IEEE}
}

@inproceedings{roumestan2021high,
  title={High-rate continuous variable quantum key distribution based on probabilistically shaped 64 and 256-QAM},
  author={Roumestan, Fran{\c{c}}ois and Ghazisaeidi, Amirhossein and Renaudier, J{\'e}r{\'e}mie and Vidarte, Luis Trigo and Diamanti, Eleni and Grangier, Philippe},
  booktitle={2021 European conference on optical communication (ECOC)},
  pages={1--4},
  year={2021},
  organization={IEEE}
}

@article{hajomer2024continuous,
  title={Continuous-variable quantum key distribution at 10 gbaud using an integrated photonic-electronic receiver},
  author={Hajomer, Adnan AE and Bruynsteen, C{\'e}dric and Derkach, Ivan and Jain, Nitin and Bomhals, Axl and Bastiaens, Sarah and Andersen, Ulrik L and Yin, Xin and Gehring, Tobias},
  journal={Optica},
  volume={11},
  number={9},
  pages={1197--1204},
  year={2024},
  publisher={Optica Publishing Group}
}

@article{milovanvcev2021high,
  title={High rate CV-QKD secured mobile WDM fronthaul for dense 5G radio networks},
  author={Milovan{\v{c}}ev, Dinka and Voki{\'c}, Nemanja and Laudenbach, Fabian and Pacher, Christoph and H{\"u}bel, Hannes and Schrenk, Bernhard},
  journal={Journal of Lightwave Technology},
  volume={39},
  number={11},
  pages={3445--3457},
  year={2021},
  publisher={OSA}
}

@article{pan2023simple,
  title={Simple and fast polarization tracking algorithm for continuous-variable quantum key distribution system using orthogonal pilot tone},
  author={Pan, Yan and Wang, Heng and Shao, Yun and Pi, Yaodi and Ye, Ting and Zhang, Shuai and Li, Yang and Huang, Wei and Xu, Bingjie},
  journal={Journal of Lightwave Technology},
  volume={41},
  number={19},
  pages={6169--6175},
  year={2023},
  publisher={IEEE}
}

@article{xu2023real,
  title={Real-time low-complexity diversity combining algorithm for free space coherent optical communication systems over atmospheric turbulence channel},
  author={Xu, Kejia and Song, Jingwei and Li, Yan and Chen, Junjie and Qiu, Jifang and Hong, Xiaobin and Guo, Hongxiang and Yang, Zhisheng and Wu, Jian},
  journal={Optics Express},
  volume={31},
  number={24},
  pages={40705--40716},
  year={2023},
  publisher={Optica Publishing Group}
}

@inproceedings{pan2024100,
  title={100 km Discrete-Modulated Continuous-Variable Quantum Key Distribution Using Probabilistic Shaped 16QAM},
  author={Pan, Yan and Wu, Mingze and Wang, Heng and Shao, Yun and Liu, Jinlu and Li, Yang and Zhang, Yichen and Huang, Wei and Xu, Bingjie},
  booktitle={2024 Asia Communications and Photonics Conference (ACP) and International Conference on Information Photonics and Optical Communications (IPOC)},
  pages={1--5},
  year={2024},
  organization={IEEE}
}

@article{qiao2024novel,
  title={Novel dispersion and timing estimation for weakly-coupled oam fiber transmission systems},
  author={Qiao, Meng and Wang, Yue and Li, Zongkai and Xiao, Yongguang and Chen, Yingyu and Li, Zhaohui and Wang, Dawei},
  journal={IEEE Photonics Technology Letters},
  volume={36},
  number={14},
  pages={913--916},
  year={2024},
  publisher={IEEE}
}

@inproceedings{dos2024real,
  title={Real-time CV-QKD Reception Resilient to Urban Atmospheric Turbulence},
  author={dos Reis Fraz{\~a}o, Jo{\~a}o and van Vliet, Vincent and van der Heide, Sjoerd and van den Hout, Menno and G{\"u}m{\"u}{\c{s}}, Kadir and Albores-Mejia, Aaron and {\v{S}}kori{\'c}, Boris and Okonkwo, Chigo},
  booktitle={CLEO: Applications and Technology},
  pages={AW3D--1},
  year={2024},
  organization={Optica Publishing Group}
}

@article{shen2023experimental,
  title={Experimental demonstration of LLO continuous-variable quantum key distribution with polarization loss compensation},
  author={Shen, Tao and Wang, Xiangyu and Chen, Ziyang and Tian, Huiping and Yu, Song and Guo, Hong},
  journal={IEEE Photonics Journal},
  volume={15},
  number={2},
  pages={1--9},
  year={2023},
  publisher={IEEE}
}

@article{chin2022digital,
  title={Digital synchronization for continuous-variable quantum key distribution},
  author={Chin, Hou-Man and Jain, Nitin and Andersen, Ulrik L and Zibar, Darko and Gehring, Tobias},
  journal={Quantum Science and Technology},
  volume={7},
  number={4},
  pages={045006},
  year={2022},
  publisher={IOP Publishing}
}

@article{freire2022neural,
  title={Neural networks-based equalizers for coherent optical transmission: Caveats and pitfalls},
  author={Freire, Pedro J and Napoli, Antonio and Spinnler, Bernhard and Costa, Nelson and Turitsyn, Sergei K and Prilepsky, Jaroslaw E},
  journal={IEEE Journal of Selected Topics in Quantum Electronics},
  volume={28},
  number={4: Mach. Learn. in Photon. Commun. and Meas. Syst.},
  pages={1--23},
  year={2022},
  publisher={IEEE}
}

@article{alshaer2024enhancing,
  title={Enhancing Performance of Continuous-Variable Quantum Key Distribution (CV-QKD) and Gaussian Modulation of Coherent States (GMCS) in Free-Space Channels under Individual Attacks with Phase-Sensitive Amplifier (PSA) and Homodyne Detection (HD)},
  author={Alshaer, Nancy and Ismail, Tawfik and Mahmoud, Haitham},
  journal={Sensors},
  volume={24},
  number={16},
  pages={5201},
  year={2024},
  publisher={MDPI}
}

@article{jin2021key,
  title={Key-sifting algorithms for continuous-variable quantum key distribution},
  author={Jin, Di and Guo, Ying and Wang, Yijun and Li, Yin and Huang, Duan},
  journal={Physical Review A},
  volume={104},
  number={1},
  pages={012616},
  year={2021},
  publisher={APS}
}

@inproceedings{anjos2023fpga,
  title={An FPGA-Based Physical Layer Approach for a CV-QKD Transmitter},
  author={Anjos, Gustavo and Almeida, Margarida and Martins, Jos{\'e} and Silva, Nuno A and Muga, Nelson J and Pinto, Armando N},
  booktitle={2023 23rd International Conference on Transparent Optical Networks (ICTON)},
  pages={1--4},
  year={2023},
  organization={IEEE}
}

@inproceedings{iqbal2024sdn,
  title={SDN-enabled continuous-variable QKD in coexistence with 8$\times$ 200 Gb/s 16-QAM classical channels},
  author={Iqbal, Masab and Villegas, Arturo and Moreolo, Michela Svaluto and Nadal, Laia and Mu{\~n}oz, Raul and Adillon, Pol and Sarmiento, Samael and Tabares, Jeison and Etcheverry, Sebastian},
  booktitle={2024 International Conference on Optical Network Design and Modeling (ONDM)},
  pages={1--3},
  year={2024},
  organization={IEEE}
}

@inproceedings{li2022quantum,
  title={Quantum key distribution post-processing: A heterogeneous computing perspective},
  author={Li, He and Wonfor, Adrian and Weerasinghe, Amanda and Alhussein, Muataz and Gong, Yupeng and Penty, Richard},
  booktitle={2022 IEEE 35th International System-on-Chip Conference (SOCC)},
  pages={1--6},
  year={2022},
  organization={IEEE}
}

@article{yan2022large,
  title={Large-scale and high-speed FPGA-based privacy amplification for quantum key distribution},
  author={Yan, Bing-Ze and Li, Qiong and Mao, Hao-Kun and Xu, Hong-Wei and Abd El-Latif, Ahmed A},
  journal={Journal of Lightwave Technology},
  volume={41},
  number={1},
  pages={169--175},
  year={2022},
  publisher={IEEE}
}

@article{yang2020high,
  title={High-speed post-processing in continuous-variable quantum key distribution based on FPGA implementation},
  author={Yang, Shen-Shen and Lu, Zhen-Guo and Li, Yong-Min},
  journal={Journal of Lightwave Technology},
  volume={38},
  number={15},
  pages={3935--3941},
  year={2020},
  publisher={IEEE}
}

@article{chen2021integrated,
  title={An integrated space-to-ground quantum communication network over 4,600 kilometres},
  author={Chen, Yu-Ao and Zhang, Qiang and Chen, Teng-Yun and Cai, Wen-Qi and Liao, Sheng-Kai and Zhang, Jun and Chen, Kai and Yin, Juan and Ren, Ji-Gang and Chen, Zhu and others},
  journal={Nature},
  volume={589},
  number={7841},
  pages={214--219},
  year={2021},
  publisher={Nature Publishing Group UK London}
}

@inproceedings{alvaro2022caramuel,
  title={Caramuel: The future of space quantum key distribution in geo},
  author={Alvaro, Angel and Pascual, Luis and Abad, Antonio and Pinto, Pedro and Alvarez-Herrero, Alberto and Belenguer, Tomas and Miravet, Carlos and Campo, Pablo and Rodriguez, Luis F and Reyes, Marcos and others},
  booktitle={2022 IEEE International Conference on Space Optical Systems and Applications (ICSOS)},
  pages={57--65},
  year={2022},
  organization={IEEE}
}

@article{gao2024optimizing,
  title={Optimizing Global Quantum Communication via Satellite Constellations},
  author={Gao, Yichen and Song, Guanqun and Zhu, Ting},
  journal={arXiv preprint arXiv:2501.00280},
  year={2024}
}

@article{cheng2025feasibility,
  title={Feasibility and parameter optimization of ground-to-satellite uplink continuous-variable quantum key distribution},
  author={Cheng, Jin and Chen, Yujie and Liu, Ao and Sun, Xin and Guo, Junjie and Yang, Bohan and Yin, Peng and Liu, Wenbo and Chen, Lanjian and Dong, Chen},
  journal={New Journal of Physics},
  volume={27},
  number={2},
  pages={023011},
  year={2025},
  publisher={IOP Publishing}
}

@article{orsucci2025assessment,
  title={Assessment of Practical Satellite Quantum Key Distribution Architectures for Current and Near-Future Missions},
  author={Orsucci, Davide and Kleinpa{\ss}, Philipp and Meister, Jaspar and De Marco, Innocenzo and H{\"a}usler, Stefanie and Strang, Thomas and Walenta, Nino and Moll, Florian},
  journal={International Journal of Satellite Communications and Networking},
  volume={43},
  number={3},
  pages={164--192},
  year={2025},
  publisher={Wiley Online Library}
}

@article{derkach2020applicability,
  title={Applicability of squeezed-and coherent-state continuous-variable quantum key distribution over satellite links},
  author={Derkach, Ivan and Usenko, Vladyslav C},
  journal={Entropy},
  volume={23},
  number={1},
  pages={55},
  year={2020},
  publisher={MDPI}
}

@article{mantey2025coexistence,
  title={On the Coexistence of Quantum and Classical Signal Transmission Over Turbulent FSO Channels},
  author={Mantey, ST and Fernandes, MA and Fernandes, GM and Silva, NA and Guiomar, FP and Monteiro, P and Pinto, AN and Muga, NJ},
  journal={Journal of Lightwave Technology},
  volume={43},
  number={3},
  pages={1043--1050},
  year={2025}
}

@article{Zhang_2019,
doi = {10.1088/2058-9565/ab19d1},
url = {https://dx.doi.org/10.1088/2058-9565/ab19d1},
year = {2019},
month = {may},
publisher = {IOP Publishing},
volume = {4},
number = {3},
pages = {035006},
author = {Zhang, Yichen and Li, Zhengyu and Chen, Ziyang and Weedbrook, Christian and Zhao, Yijia and Wang, Xiangyu and Huang, Yundi and Xu, Chunchao and Zhang, Xiaoxiong and Wang, Zhenya and Li, Mei and Zhang, Xueying and Zheng, Ziyong and Chu, Binjie and Gao, Xinyu and Meng, Nan and Cai, Weiwen and Wang, Zheng and Wang, Gan and Yu, Song and Guo, Hong},
title = {Continuous-variable QKD over 50 km commercial fiber},
journal = {Quantum Science and Technology}
}

@article{Zhang2019PIC,
  author  = {Zhang, G. and Haw, J. Y. and Cai, H. and Xu, F. and Assad, S. M. and Fitzsimons, J. F. and Zhou, X. and Zhang, Y. and Yu, S. and Wu, J. and Ser, W. and Kwek, L. C. and Liu, A. Q.},
  title   = {An integrated silicon photonic chip platform for continuous-variable quantum key distribution},
  journal = {Nature Photonics},
  year    = {2019},
  volume  = {13},
  number  = {12},
  pages   = {839--842},
  month   = {dec},
  doi     = {10.1038/s41566-019-0504-5},
  url     = {https://doi.org/10.1038/s41566-019-0504-5}
}

@ARTICLE{Yang2020HighSpeed,
  author={Yang, Shen-Shen and Lu, Zhen-Guo and Li, Yong-Min},
  journal={Journal of Lightwave Technology}, 
  title={High-Speed Post-Processing in Continuous-Variable Quantum Key Distribution Based on FPGA Implementation}, 
  year={2020},
  volume={38},
  number={15},
  pages={3935-3941},
  doi={10.1109/JLT.2020.2985408}}

@article{Pirandola_2020,
author = {S. Pirandola and U. L. Andersen and L. Banchi and M. Berta and D. Bunandar and R. Colbeck and D. Englund and T. Gehring and C. Lupo and C. Ottaviani and J. L. Pereira and M. Razavi and J. Shamsul Shaari and M. Tomamichel and V. C. Usenko and G. Vallone and P. Villoresi and P. Wallden},
journal = {Adv. Opt. Photon.},
number = {4},
pages = {1012--1236},
publisher = {Optica Publishing Group},
title = {Advances in quantum cryptography},
volume = {12},
month = {Dec},
year = {2020},
url = {https://opg.optica.org/aop/abstract.cfm?URI=aop-12-4-1012},
doi = {10.1364/AOP.361502}
}

@article{Weerasinghe2023,
  author  = {Weerasinghe, Amanda and Alhussein, Muataz and Alderton, Adam and Wonfor, Adrian and Penty, Richard},
  title   = {Practical, high-speed {G}aussian coherent state continuous variable quantum key distribution with real-time parameter monitoring, optimised slicing, and post-processed key distillation},
  journal = {Scientific Reports},
  year    = {2023},
  volume  = {13},
  number  = {1},
  pages   = {21543},
  month   = {dec},
  doi     = {10.1038/s41598-023-47517-7},
  url     = {https://doi.org/10.1038/s41598-023-47517-7}
}

@misc{ng2025gigabitqkd,
      title={Gigabit-rate Quantum Key Distribution on Integrated Photonic Chips}, 
      author={Si Qi Ng and Florian Kanitschar and Gong Zhang and Chao Wang},
      year={2025},
      eprint={2504.08298},
      archivePrefix={arXiv},
      primaryClass={quant-ph},
      url={https://arxiv.org/abs/2504.08298}
}

@INPROCEEDINGS{Sun2024HighSpeedFPGA,
  author={Sun, Jialu and Cheng, Hao and Jin, Zeyuan and Chen, Yong and Chen, Genlong and Ren, Lihong and Jiang, Xueqin},
  booktitle={2024 9th International Conference on Communication, Image and Signal Processing (CCISP)}, 
  title={A High-Speed FPGA Implementation of the Multidimensional Reconciliation for Continuous Variables Quantum Key Distribution}, 
  year={2024},
  volume={},
  number={},
  pages={170-174},
  doi={10.1109/CCISP63826.2024.10765578}}

@phdthesis{WatermanEECS2016_1,
    Author= {Waterman, Andrew},
    Title= {Design of the RISC-V Instruction Set Architecture},
    School= {EECS Department, University of California, Berkeley},
    Year= {2016},
    Month= {Jan},
    Url= {http://www2.eecs.berkeley.edu/Pubs/TechRpts/2016/EECS-2016-1.html},
    Number= {UCB/EECS-2016-1}
}

@inproceedings{Dorflinger2021RISCV,
author = {D\"{o}rflinger, Alexander and Albers, Mark and Kleinbeck, Benedikt and Guan, Yejun and Michalik, Harald and Klink, Raphael and Blochwitz, Christopher and Nechi, Anouar and Berekovic, Mladen},
title = {A comparative survey of open-source application-class RISC-V processor implementations},
year = {2021},
isbn = {9781450384049},
publisher = {Association for Computing Machinery},
address = {New York, NY, USA},
url = {https://doi.org/10.1145/3457388.3458657},
doi = {10.1145/3457388.3458657},
booktitle = {Proceedings of the 18th ACM International Conference on Computing Frontiers},
pages = {12–20},
numpages = {9},
location = {Virtual Event, Italy},
series = {CF '21}
}

@inproceedings{Kreinberg2020,
  author={Kreinberg, Sören and Koltchanov, Igor and Novik, Piotr and Alreesh, Saleem and Laudenbach, Fabian and Pacher, Christoph and Hübel, Hannes and Richter, André},
  booktitle={2019 21st International Conference on Transparent Optical Networks (ICTON)}, 
  title={Modelling Weak-Coherent CV-QKD Systems Using a Classical Simulation Framework}, 
  year={2019},
  volume={},
  number={},
  pages={1-4},
  doi={10.1109/ICTON.2019.8840253}}

@inproceedings{Pietri20242024,
  author    = {Yoann Piétri and Matteo Schiavon and Valentina Marulanda Acosta and Baptiste Gouraud and Luis Trigo Vidarte and Philippe Grangier and Amine Rhouni and Eleni Diamanti},
  title     = {QOSST: A Highly Modular Open Source Platform for Continuous Variable Quantum Key Distribution Applications},
  booktitle = {Quantum 2.0 Conference and Exhibition, Technical Digest Series},
  year      = {2024},
  publisher = {Optica Publishing Group},
  address   = {Rotterdam, Netherlands},
  doi       = {10.1364/QUANTUM.2024.QTh4B.4},
  paper     = {QTh4B.4}
}

@misc{galvao2025nnexcessnoise,
      title={Neural network for excess noise estimation in continuous-variable quantum key distribution under composable finite-size security}, 
      author={Lucas Q. Galvão and Davi Juvêncio G. de Sousa and Micael Andrade Dias and Nelson Alves Ferreira Neto},
      year={2025},
      eprint={2507.23117},
      archivePrefix={arXiv},
      primaryClass={quant-ph},
      url={https://arxiv.org/abs/2507.23117}
}

@article{yang2023information,
  title={Information reconciliation of continuous-variables quantum key distribution: principles, implementations and applications},
  author={Yang, S. and Yan, Z. and Yang, H. and et al.},
  journal={EPJ Quantum Technology},
  volume={10},
  pages={40},
  year={2023},
  doi={10.1140/epjqt/s40507-023-00197-8}
}

@article{milicevic2018quasi,
  title={Quasi-cyclic multi-edge LDPC codes for long-distance quantum cryptography},
  author={Milicevic, M. and Feng, C. and Zhang, L.M. and et al.},
  journal={npj Quantum Information},
  volume={4},
  pages={21},
  year={2018},
  publisher={Nature Publishing Group},
  doi={10.1038/s41534-018-0070-6}
}

@article{PhysRevA.81.062343,
  title = {Finite-size analysis of a continuous-variable quantum key distribution},
  author = {Leverrier, Anthony and Grosshans, Fr\'ed\'eric and Grangier, Philippe},
  journal = {Phys. Rev. A},
  volume = {81},
  issue = {6},
  pages = {062343},
  numpages = {11},
  year = {2010},
  month = {Jun},
  publisher = {American Physical Society},
  doi = {10.1103/PhysRevA.81.062343},
  url = {https://link.aps.org/doi/10.1103/PhysRevA.81.062343}
}

@misc{usenko2025continuousvariablequantumcommunication,
      title={Continuous-variable quantum communication}, 
      author={Vladyslav C. Usenko and Antonio Acín and Romain Alléaume and Ulrik L. Andersen and Eleni Diamanti and Tobias Gehring and Adnan A. E. Hajomer and Florian Kanitschar and Christoph Pacher and Stefano Pirandola and Valerio Pruneri},
      year={2025},
      eprint={2501.12801},
      archivePrefix={arXiv},
      primaryClass={quant-ph},
      url={https://arxiv.org/abs/2501.12801}, 
}

@Article{Diamanti2015,
AUTHOR = {Diamanti, Eleni and Leverrier, Anthony},
TITLE = {Distributing Secret Keys with Quantum Continuous Variables: Principle, Security and Implementations},
JOURNAL = {Entropy},
VOLUME = {17},
YEAR = {2015},
NUMBER = {9},
PAGES = {6072--6092},
url = {https://www.mdpi.com/1099-4300/17/9/6072},
ISSN = {1099-4300},
DOI = {10.3390/e17096072}
}

@article{Leverrier_2008,
   title={Multidimensional reconciliation for a continuous-variable quantum key distribution},
   volume={77},
   ISSN={1094-1622},
   url={http://dx.doi.org/10.1103/PhysRevA.77.042325},
   DOI={10.1103/physreva.77.042325},
   number={4},
   journal={Physical Review A},
   publisher={American Physical Society (APS)},
   author={Leverrier, Anthony and Alléaume, Romain and Boutros, Joseph and Zémor, Gilles and Grangier, Philippe},
   year={2008},
   month=apr }

@book{moon2005error,
  author    = {Todd K. Moon},
  title     = {Error Correction Coding: Mathematical Methods and Algorithms},
  publisher = {Wiley-Interscience},
  address   = {Hoboken, NJ},
  year      = {2005},
  edition   = {1},
  isbn      = {978-0-471-64800-0},
  doi       = {10.1002/0471739219}
}

@article{Multi-edge-LDPCcodes,
author = {Richardson, Tom and Urbanke, Rüdiger},
year = {2002},
month = {01},
pages = {},
title = {Multi-edge type LDPC codes},
journal = {ISIT talk}
}

@article{li2020high,
  title={High-throughput GPU layered decoder of quasi-cyclic multi-edge type low density parity check codes in continuous-variable quantum key distribution systems},
  author={Li, Yang and Zhang, Xiaofang and Li, Yong and Xu, Bingjie and Ma, Li and Yang, Jie and Huang, Wei},
  journal={Scientific reports},
  volume={10},
  number={1},
  pages={14561},
  year={2020},
  publisher={Nature Publishing Group UK London}
}

@article{doi:10.1126/sciadv.adi9474,
author = {Adnan A. E. Hajomer  and Ivan Derkach  and Nitin Jain  and Hou-Man Chin  and Ulrik L. Andersen  and Tobias Gehring },
title = {Long-distance continuous-variable quantum key distribution over 100-km fiber with local local oscillator},
journal = {Science Advances},
volume = {10},
number = {1},
pages = {eadi9474},
year = {2024},
doi = {10.1126/sciadv.adi9474}}

@Article{Huang2016,
author={Huang, Duan
and Huang, Peng
and Lin, Dakai
and Zeng, Guihua},
title={Long-distance continuous-variable quantum key distribution by controlling excess noise},
journal={Scientific Reports},
year={2016},
month={Jan},
day={13},
volume={6},
number={1},
pages={19201},
issn={2045-2322},
doi={10.1038/srep19201},
url={https://doi.org/10.1038/srep19201}
}

@article{Fossier_2009,
doi = {10.1088/1367-2630/11/4/045023},
url = {https://dx.doi.org/10.1088/1367-2630/11/4/045023},
year = {2009},
month = {apr},
publisher = {},
volume = {11},
number = {4},
pages = {045023},
author = {Fossier, S and Diamanti, E and Debuisschert, T and Villing, A and Tualle-Brouri, R and Grangier, P},
title = {Field test of a continuous-variable quantum key distribution prototype},
journal = {New Journal of Physics}
}

@Article{Jouguet2013,
author={Jouguet, Paul
and Kunz-Jacques, S{\'e}bastien
and Leverrier, Anthony
and Grangier, Philippe
and Diamanti, Eleni},
title={Experimental demonstration of long-distance continuous-variable quantum key distribution},
journal={Nature Photonics},
year={2013},
month={May},
day={01},
volume={7},
number={5},
pages={378-381},
issn={1749-4893},
doi={10.1038/nphoton.2013.63},
url={https://doi.org/10.1038/nphoton.2013.63}
}

@article{Jouguet:12,
author = {Paul Jouguet and S\'{e}bastien Kunz-Jacques and Thierry Debuisschert and Simon Fossier and Eleni Diamanti and Romain All\'{e}aume and Rosa Tualle-Brouri and Philippe Grangier and Anthony Leverrier and Philippe Pache and Philippe Painchault},
journal = {Opt. Express},
number = {13},
pages = {14030--14041},
publisher = {Optica Publishing Group},
title = {Field test of classical symmetric encryption with continuous variables quantum key distribution},
volume = {20},
month = {Jun},
year = {2012},
url = {https://opg.optica.org/oe/abstract.cfm?URI=oe-20-13-14030},
doi = {10.1364/OE.20.014030},
}

@article{PhysRevApplied.13.024058,
  title = {One-Time Shot-Noise Unit Calibration Method for Continuous-Variable Quantum Key Distribution},
  author = {Zhang, Yichen and Huang, Yundi and Chen, Ziyang and Li, Zhengyu and Yu, Song and Guo, Hong},
  journal = {Phys. Rev. Appl.},
  volume = {13},
  issue = {2},
  pages = {024058},
  numpages = {17},
  year = {2020},
  month = {Feb},
  publisher = {American Physical Society},
  doi = {10.1103/PhysRevApplied.13.024058},
  url = {https://link.aps.org/doi/10.1103/PhysRevApplied.13.024058}
}

@ARTICLE{Wang,
  author={Wang, Xu-Yang and Guo, Xu-Bo and Jia, Yan-Xiang and Zhang, Yu and Lu, Zhen-Guo and Liu, Jian-Qiang and Li, Yong-Min},
  journal={Journal of Lightwave Technology}, 
  title={Accurate Shot-Noise-Limited Calibration of a Time-Domain Balanced Homodyne Detector for Continuous-Variable Quantum Key Distribution}, 
  year={2023},
  volume={41},
  number={17},
  pages={5518-5528},
  doi={10.1109/JLT.2023.3264234}}

@article{PhysRevResearch.3.043014,
  title = {Composable security for continuous variable quantum key distribution: Trust levels and practical key rates in wired and wireless networks},
  author = {Pirandola, Stefano},
  journal = {Phys. Rev. Res.},
  volume = {3},
  issue = {4},
  pages = {043014},
  numpages = {18},
  year = {2021},
  month = {Oct},
  publisher = {American Physical Society},
  doi = {10.1103/PhysRevResearch.3.043014},
  url = {https://link.aps.org/doi/10.1103/PhysRevResearch.3.043014}
}

@book{casella2024statistical,
  author    = {George Casella and Roger W. Berger},
  title     = {Statistical Inference},
  series    = {CRC Texts in Statistical Science},
  publisher = {CRC Press},
  address   = {Boca Raton},
  year      = {2024},
  edition   = {2},
  isbn      = {978-1-003-45628-5},
  doi       = {10.1201/9781003456285}
}

@article{PhysRevA.93.042343,
  title = {Estimation of output-channel noise for continuous-variable quantum key distribution},
  author = {Thearle, Oliver and Assad, Syed M. and Symul, Thomas},
  journal = {Phys. Rev. A},
  volume = {93},
  issue = {4},
  pages = {042343},
  numpages = {6},
  year = {2016},
  month = {Apr},
  publisher = {American Physical Society},
  doi = {10.1103/PhysRevA.93.042343},
  url = {https://link.aps.org/doi/10.1103/PhysRevA.93.042343}
}

@article{PhysRevResearch.6.023321,
  title = {Improved composable key rates for CV-QKD},
  author = {Pirandola, Stefano and Papanastasiou, Panagiotis},
  journal = {Phys. Rev. Res.},
  volume = {6},
  issue = {2},
  pages = {023321},
  numpages = {13},
  year = {2024},
  month = {Jun},
  publisher = {American Physical Society},
  doi = {10.1103/PhysRevResearch.6.023321},
  url = {https://link.aps.org/doi/10.1103/PhysRevResearch.6.023321}
}

@article{PhysRevLett.125.010502,
  title = {Long-Distance Continuous-Variable Quantum Key Distribution over 202.81 km of Fiber},
  author = {Zhang, Yichen and Chen, Ziyang and Pirandola, Stefano and Wang, Xiangyu and Zhou, Chao and Chu, Binjie and Zhao, Yijia and Xu, Bingjie and Yu, Song and Guo, Hong},
  journal = {Phys. Rev. Lett.},
  volume = {125},
  issue = {1},
  pages = {010502},
  numpages = {6},
  year = {2020},
  month = {Jun},
  publisher = {American Physical Society},
  doi = {10.1103/PhysRevLett.125.010502},
  url = {https://link.aps.org/doi/10.1103/PhysRevLett.125.010502}
}

@article{Cai_2009,
    doi = {10.1088/1367-2630/11/4/045024},
    url = {https://dx.doi.org/10.1088/1367-2630/11/4/045024},
    year = {2009},
    month = {apr},
    publisher = {IOP Publishing},
    volume = {11},
    number = {4},
    pages = {045024},
    author = {Cai, Raymond Y Q and Scarani, Valerio},
    title = {Finite-key analysis for practical implementations of quantum key distribution},
    journal = {New Journal of Physics}
}

@article{PhysRevA.97.022316,
  title = {Integrating machine learning to achieve an automatic parameter prediction for practical continuous-variable quantum key distribution},
  author = {Liu, Weiqi and Huang, Peng and Peng, Jinye and Fan, Jianping and Zeng, Guihua},
  journal = {Phys. Rev. A},
  volume = {97},
  issue = {2},
  pages = {022316},
  numpages = {9},
  year = {2018},
  month = {Feb},
  publisher = {American Physical Society},
  doi = {10.1103/PhysRevA.97.022316},
  url = {https://link.aps.org/doi/10.1103/PhysRevA.97.022316}
}

@article{Luo_2022,
doi = {10.1088/1674-1056/ac2807},
url = {https://dx.doi.org/10.1088/1674-1056/ac2807},
year = {2022},
month = {feb},
publisher = {Chinese Physical Society and IOP Publishing Ltd},
volume = {31},
number = {2},
pages = {020306},
author = {Luo, Hao and Wang, Yi-Jun and Ye, Wei and Zhong, Hai and Mao, Yi-Yu and Guo, Ying},
title = {Parameter estimation of continuous variable quantum key distribution system via artificial neural networks},
journal = {Chinese Physics B}
}

@article{gg02,
  title = {Continuous Variable Quantum Cryptography Using Coherent States},
  author = {Grosshans, Fr\'ed\'eric and Grangier, Philippe},
  journal = {Phys. Rev. Lett.},
  volume = {88},
  issue = {5},
  pages = {057902},
  numpages = {4},
  year = {2002},
  month = {Jan},
  publisher = {American Physical Society},
  doi = {10.1103/PhysRevLett.88.057902},
  url = {https://link.aps.org/doi/10.1103/PhysRevLett.88.057902}
}

@article{Furrer100502,
  title={Continuous variable quantum key distribution: finite-key analysis of composable security against coherent attacks},
  author={Furrer, Fabian and Franz, Torsten and Berta, Mario and Leverrier, Anthony and Scholz, Volkher B and Tomamichel, Marco and Werner, Reinhard F},
  journal={Physical review letters},
  volume={109},
  number={10},
  pages={100502},
  year={2012},
  publisher={APS}
}

@ARTICLE{Konig2009,
  author={Konig, Robert and Renner, Renato and Schaffner, Christian},
  journal={IEEE Transactions on Information Theory}, 
  title={The Operational Meaning of Min- and Max-Entropy}, 
  year={2009},
  volume={55},
  number={9},
  pages={4337-4347},
  doi={10.1109/TIT.2009.2025545}}

@book{de2021digital,
  author    = {Darli Augusto de Arruda Mello and Fabio Aparecido Barbosa},
  title     = {Digital Coherent Optical Systems: Architecture and Algorithms},
  publisher = {Springer},
  address   = {Cham},
  year      = {2021},
  edition   = {1},
  isbn      = {978-3-030-66541-8},
  doi       = {10.1007/978-3-030-66541-8},
  url       = {https://doi.org/10.1007/978-3-030-66541-8}
}

@ARTICLE{9927353,
  author={Pereira, Daniel and Pinto, Armando N. and Silva, Nuno A.},
  journal={Journal of Lightwave Technology}, 
  title={Polarization Diverse True Heterodyne Receiver Architecture for Continuous Variable Quantum Key Distribution}, 
  year={2023},
  volume={41},
  number={2},
  pages={432-439},
  doi={10.1109/JLT.2022.3216754}}

\end{document}